\documentclass[a4paper,11pt]{article}
\usepackage{pub}
\usepackage{braket}
\usepackage{xcolor}
\usepackage{hyperref}
\usepackage{fontawesome5}

\usepackage{booktabs}

\usepackage{placeins}
\usepackage{float}

\makeatletter
\newcommand\blfootnote[1]{%
  \begingroup
  \renewcommand\thefootnote{}\footnote{#1}%
  \addtocounter{footnote}{-1}%
  \endgroup
}
\makeatother

\title{\boldmath KineticXGPU: A Tensorized Collision Operator for Dark-Sector Self-Scattering}

\author{Esau Cervantes}

\affiliation{National Centre for Nuclear Research,\\
Pasteura 7, 02-093 Warsaw, Poland}

\emailAdd{esau.cervantes@ncbj.gov.pl}

\abstract{
We present KineticXGPU, a PyTorch-based implementation of the \(2\to2\) elastic self-collision operator for dark-sector momentum distributions. The discretized collision operator can be expressed as tensor contractions and is therefore well suited for GPUs. As an application, we study a two-source freeze-in scenario in which the final distribution can develop a bimodal shape. We show that increasing the strength of elastic self-interactions progressively erases this structure and drives the distribution toward a Maxwell--Boltzmann distribution. We find that $2\to2$ self-interactions can reduce the warmness of nonthermal frozen-in dark matter, illustrating how they may alleviate structure-formation constraints in late-decay scenarios. We compare the phase-space formulation with a set of fluid equations that couple the number density and velocity dispersion. We also compare CPU and GPU runtimes and demonstrate the computational advantage of the tensorized approach. The code is publicly available on GitHub~\href{https://github.com/EsauCervantes/KineticXGPU}{\textcolor{black}{\faGithub{}}}.

\blfootnote{
\footnotesize
\href{https://github.com/EsauCervantes/KineticXGPU}
{\textcolor{black}{\faGithub\ \nolinkurl{https://github.com/EsauCervantes/KineticXGPU}}}
}

}

\begin{document}
\maketitle
\flushbottom

\section{Introduction}
\label{sec:intro}

The existence of dark matter (DM) is well established by a wide range of cosmological and astrophysical observations, but its microscopic nature remains unknown~\cite{Bertone:2010zza,Cirelli:2024ssz}. Although weakly interacting massive particles (WIMPs) and thermal freeze-out remain viable possibilities, the absence of a confirmed signal in direct-detection, indirect-detection, and collider searches (see e.g.~\cite{Arcadi:2024ukq,LZ:2024zvo,LZ:2025igz,XENON:2025vwd,ATLAS:2024kpy}) has motivated the study of alternative production histories. Among them, scenarios in which DM is produced or evolves out of equilibrium include freeze-in, production from late decays, and freeze-out with early kinetic decoupling or velocity-dependent annihilation~\cite{Hall:2009bx,Feng:2010zp,Binder:2017rgn}. In particular, freeze-in provides a minimal framework in which the DM abundance is generated gradually through rare decays or annihilations in the early Universe~\cite{Hall:2009bx}. The particles sourcing DM may belong to either the Standard Model (SM) thermal bath or a more extended dark sector. Since the interactions responsible for production are feeble, the DM population does not thermalize with the visible sector and its phase-space distribution can retain information about the time and kinematics of production.

Irrespective of the production mechanism, DM may also possess sizable self-interactions, which have been proposed as a possible way to modify halo structure and address small-scale tensions in collisionless cold dark matter~\cite{Flores:1994gz,Moore:1994yx,2011AJ....142...24O,2011ApJ...742...20W,2011MNRAS.415L..40B}. In the early Universe, they can also play a central role in determining the relic abundance. If self-number-changing reactions are efficient, the dark sector can reach chemical equilibrium and later undergo chemical decoupling, leaving a finite comoving abundance. This is the mechanism behind cannibal dark matter, where the rest mass is converted into kinetic energy during self-number-changing freeze-out~\cite{Carlson:1992fn,deLaix:1995vi,Hochberg:2014dra,Buen-Abad:2018mas,Chatterjee:2019olf,Erickcek:2020wzd,Heimersheim:2020aoc,Hufnagel:2022aiz}. Similar dynamics can arise when a dark-sector population is first generated by freeze-in and subsequently evolves through self-interactions. In that case, reactions such as $2\to3$ or $2\to4$ can convert the excess kinetic energy of the initially produced particles into additional dark-sector particles, driving the system toward chemical equilibrium~\cite{Heeba:2018wtf,March-Russell:2020nun,Bernal:2015xba,Cervantes:2024ipg,Bernal:2025osg}. In fact, this process typically occurs once the dark sector has already reached local thermal equilibrium (LTE), meaning a Bose--Einstein (for bosons) or Fermi--Dirac (for fermions) shape, or, in the case of a dilute or non-relativistic fluid, a Maxwell--Boltzmann shape. The reason is that elastic \(2\to2\) scatterings are usually less phase-space suppressed and arise at a lower order in the couplings than self-number-changing reactions. They, therefore, tend to establish LTE first, after which chemical equilibration can be described in terms of the DM temperature and chemical potential. However, this hierarchy of rates does not guarantee that LTE is reached in all viable regions of parameter space.

The efficiency of kinetic equilibration is controlled by the elastic self-scattering rate, whose underlying cross section is constrained by astrophysical observations, such as those of the Bullet Cluster~\cite{Randall:2008ppe}. The strength of the self-interactions may then be insufficient to enforce LTE in the early Universe, while still modifying the phase-space distribution without erasing the nonthermal features. This can occur, for instance, when DM is sourced by several states or production channels operating at different times and injecting particles with different momenta~\cite{Hryczuk:2022gay,Chatterjee:2025vdz,Chatterjee:2026mkh,Du:2021jcj}. In such cases, the momentum distribution is not merely a technical detail; it can lead to appreciable changes in the small-scale matter power spectrum, which depends on the dark matter velocity distribution and its free-streaming history, and can thus be used to place constraints on warm dark matter (WDM)~\cite{Murgia:2018now,DEramo:2020gpr,Dienes:2020bmn,Decant:2021mhj,Dienes:2021cxp,DEramo:2025jsb,Zhao:2026wxi,Dienes:2026prl,Thoss:2026slt}. These effects are therefore subject to constraints from the Lyman-$\alpha$ forest~\cite{McQuinn:2015icp} because a nonthermal distribution can suppress structure formation differently from a thermal one with the same relic abundance~\cite{Huo:2019bjf,Dvorkin:2020xga,Garani:2022yzj}; and, even if full LTE is not achieved, self-scattering can still partially distort, smooth, or erase these features as the dark sector attempts to equilibrate. 

In fact, the impact of dark-matter self-interactions on structure formation has been studied through complementary approaches in refs.~\cite{Egana-Ugrinovic:2021gnu,Yunis:2020woq,Yunis:2021fgz,Garani:2022yzj}, which compute modifications of the linear matter power spectrum and derive or estimate Lyman-$\alpha$ constraints. These analyses nevertheless treat the homogeneous distribution through equilibrium or collisionless limits using fluid, tight-coupling, or relaxation-time approximations where appropriate for the perturbations. In particular, refs.~\cite{Yunis:2021fgz,Garani:2022yzj} identify a full time-dependent treatment of the intermediate background evolution as a missing refinement. Thus, to the best of our knowledge, existing Lyman-$\alpha$ analyses do not explicitly evolve an arbitrary nonthermal homogeneous distribution through partial relaxation using the full $2\to 2$ collision operator. Relatedly, freeze-in phase-space information can also be relevant for other cosmological observables, such as dark matter isocurvature perturbations~\cite{Heikinheimo:2016yds,Bellomo:2022qbx}. Quantifying this requires following the evolution of the full phase-space distribution.

This motivates a momentum-dependent solution of the Boltzmann equation that includes elastic \(2\to2\) self-scattering. Since this process is expected to be more efficient than self-number-changing reactions, the corresponding collision operator is sufficient for studying relaxation toward equilibrium. In practice, the solution of the Boltzmann equation is numerically demanding because the collision operator couples different regions of phase-space; i.e., after discretization, the elastic self-collision term becomes a sum over the momentum grid. This computational difficulty may be mitigated by the fact that the resulting structure can be reorganized into repeated tensor contractions because the dominant operations are parallelizable and therefore well suited for GPUs.

Several public tools already address dark matter evolution at the phase-space level. For example, DRAKE implements fluid and phase-space Boltzmann solvers for dark matter production~\cite{Binder:2021bmg}. More recently, BEST was introduced as a Python framework for self-interaction collision integrals~\cite{Yoon:2026rce}. These tools provide robust numerical treatments of dark matter production and thermalization, but public phase-space Boltzmann solvers in this context remain largely based on CPU libraries. On the other hand, GPU-based solvers have been explored in related scenarios, such as Einstein--Boltzmann solvers used for cosmic microwave background (CMB) and large-scale-structure calculations. Examples include GPU versions of CAMB and differentiable solvers written in JAX~\cite{gCAMB,DISCODJ,ABCMB}. These codes evolve cosmological perturbations and compute observables such as CMB spectra or matter power spectra. In the present work, we focus on the unperturbed dark-sector distribution function and the self-collision operator governing its relaxation toward LTE, while the computational reason for using GPUs remains the same. The discretized collision operator involves many repeated operations over a momentum grid, and its structure can be directly mapped onto parallelizable tensor operations.

The present code addresses this gap by explicitly demonstrating at the phase-space level how thermalization interpolates continuously between collisionless and LTE distributions relevant for Lyman-$\alpha$ constraints. As an application, we consider a scalar contact interaction in which DM is produced via two freeze-in sources on different timescales. In the absence of self-interactions, the final distribution contains two separated momentum components. We track the comoving warmness and compare the tensorial output with the fluid treatment in which only the number density and temperature are evolved.

The paper is organized as follows: first, we review the phase-space Boltzmann equation for freeze-in production and elastic self-scattering; secondly, we introduce the particle-physics scenario used in the numerical study; thirdly, after describing the discretization of the collision operator and the numerical implementation, we present the cosmological phase-space evolution and discuss its physical implications. In the final section, we present our discussion and conclusions.

\section{Boltzmann equation}
\label{sec:boltzmann}

We consider a dark matter species $\chi$ produced via the freeze-in mechanism in a homogeneous and expanding background. Its phase-space distribution evolves according to the full Boltzmann equation (fBE),
\begin{equation}
\label{eq:fbe}
	\left( \partial_t - H p\,\partial_p \right) f(t,p)
	= C_\text{el}[f] + C_{\rm FI}(t,p) \,,
\end{equation}
where $p$ is the physical momentum, $H$ is the Hubble rate, $C_{\rm FI}$ denotes the freeze-in source term, and $C_\text{el}$ is the elastic $2\to2$ self-collision operator.\footnote{Here $C_\text{el}$ denotes the elastic self-scattering operator for $\chi\chi\leftrightarrow\chi\chi$. In other contexts, similar notation is also used for elastic scattering between DM and an external bath species, such as SM particles~\cite{Binder:2021bmg}; this is not the process considered here.} The freeze-in term injects dark matter particles into phase space, while the elastic term only redistributes energy and momentum within the dark sector. 

For a generic two-body decay source, with $A$ as the parent particle and $\chi$ and $X$ as the daughter particles, the contribution from $A\to\chi X$ to the collision term for a $\chi$ particle with momentum $p_1$ can be written as
\begin{equation}
\label{eq:CFI_decay_general}
C_{\rm FI}^{A\to\chi X}(t,p_1)
=
\frac{1}{2E_1g_\chi}
\int d\Pi_A\,d\Pi_X\,
(2\pi)^4\delta^{(4)}(p_A-p_1-p_X)\,
|\mathcal{M}_{A\to\chi X}|^2\,
f_A(t,p_A) \,,
\end{equation}
where $g_\chi$ is the number of internal degrees of freedom of $\chi$, while
\begin{equation}
	d\Pi_i \equiv \frac{d^3p_i}{(2\pi)^3 2E_i} \, ,
	\qquad
	E_i = \sqrt{p_i^2+m_i^2} \,.
\end{equation}
Inverse processes and final-state statistical factors have been neglected. In the applications considered below, $X=\chi$. For the elastic self-scattering term, we write the process as $\chi_1\chi_2\leftrightarrow \chi_3\chi_4$, where the subscripts label the momenta. The collision operator is
\begin{equation}
\label{eq:C22_general}
C_\text{el}[f_1]
=
\frac{1}{2E_1g_\chi}
\int d\Pi_2\,d\Pi_3\,d\Pi_4\,
(2\pi)^4\delta^{(4)}(p_1+p_2-p_3-p_4)\,
|\mathcal{M}|^2\,
\Lambda_{12\leftrightarrow 34}[f_1] \,,
\end{equation}
where $f_i\equiv f(t,p_i)$. The squared matrix elements are summed over internal degrees of freedom. The statistical factor $\Lambda_{12\leftrightarrow 34}[f_1]$ is defined as
\begin{equation}
\label{eq:Lambda_quantum}
\Lambda_{12\leftrightarrow 34}[f_1]
\equiv
f_3 f_4 (1+\eta_1 f_1)(1+\eta_2 f_2)
-
f_1 f_2 (1+\eta_3 f_3)(1+\eta_4 f_4) \, .
\end{equation}
The parameters $\eta_i$ encode the quantum statistics of particle $i$, and are +1 for bosons and -1 for fermions. In the dilute limit $f_i\ll 1$, these effects can be neglected,
\begin{equation}
\Lambda_{12\leftrightarrow 34}[f_1]
\simeq
f_3 f_4 - f_1 f_2 \, .
\end{equation}

The elastic collision operator conserves particle number and energy inside the dark sector,
\begin{equation}
\label{eq:C22_conservation}
	g_\chi\int_p C_\text{el}[f] = 0,
	\qquad
	g_\chi\int_p E\,C_\text{el}[f] = 0 ,
\end{equation}
where $\int_p \equiv \int \frac{d^3p}{(2\pi)^3}$, while the moments of the freeze-in source are
\begin{align}
	C_0^{\rm FI}(t)
	&=
	g_\chi\int_p C_{\rm FI}(t,p) ,
	\qquad
	C_E^{\rm FI}(t)
	=
	g_\chi\int_p E\,C_{\rm FI}(t,p) .
\end{align}
Taking the zeroth and energy moments of eq.~\eqref{eq:fbe} gives
\begin{align}
\label{eq:number_moment}
	\dot n_\chi + 3H n_\chi &= C_0^{\rm FI} ,
	\\
\label{eq:energy_moment}
	\dot \rho_\chi + 3H(\rho_\chi+P_\chi) &= C_E^{\rm FI} \,,
\end{align}
where $n_\chi$, $\rho_\chi$ and $P_\chi$ are the number density, energy density and pressure of the DM, respectively. The moment equations track
the total abundance and energy injection.

For comparison with the fBE solution, we solve a coupled Boltzmann system in which the dark matter distribution is assumed to have a thermal shape with time-dependent temperature and chemical potential; i.e., we take $f = z\,e^{-E/T_\chi}$, where $z$ is the fugacity and $T_\chi$ the DM temperature. Using the scale factor as the independent time variable, the set of coupled Boltzmann equations (cBE) is
\begin{subequations}
\label{eq:cBE}
\begin{align}
        \frac{dY_\chi}{da}
        &=
        C_0^{\rm FI}/(aHs),
        \label{eq:cBE_Y_a}
        \\
        \frac{dT_\chi}{da}
        &=
        \frac{
        C_E^{\rm FI}/(aH)
        -
        \dfrac{3}{a}\left(\rho_\chi+P_\chi\right)
        -
        \dfrac{\partial \rho_\chi}{\partial Y_\chi}
        \dfrac{dY_\chi}{da}
        -
        \dfrac{\partial \rho_\chi}{\partial a}
        }{
        \dfrac{\partial \rho_\chi}{\partial T_\chi}
        } ,
        \label{eq:cBE_T_a}
\end{align}
\end{subequations}
where $Y_\chi=n_\chi/s$ and $s$ is the entropy density of the SM bath. The partial derivatives in eq.~\eqref{eq:cBE_T_a} are evaluated by regarding \(\rho_\chi=\rho_\chi(Y_\chi,T_\chi,a)\). With the Maxwell--Boltzmann ansatz, the system is closed,
\begin{equation}
\label{eq:rho_P_MB}
\begin{aligned}
        \rho_\chi
        &=
        \frac{g_\chi}{2\pi^2}\,
        z\,m_\chi^3 T_\chi
        \left[
        K_1\!\left(\frac{m_\chi}{T_\chi}\right)
        +
        3\,\frac{T_\chi}{m_\chi}\,
        K_2\!\left(\frac{m_\chi}{T_\chi}\right)
        \right],
        \\[1ex]
        P_\chi
        &=n_\chi T_\chi=
        \frac{g_\chi}{2\pi^2}\,
        z\,m_\chi^2 T_\chi^2\,
        K_2\!\left(\frac{m_\chi}{T_\chi}\right)\,,
\end{aligned}
\end{equation}
together with \(z=Y_\chi/Y_\chi^{\rm eq}\), where $Y_\chi^\text{eq}=n_\chi^\text{eq}/s$. Here $K_n$ is the modified Bessel function of the second kind and order $n$. 

Finally, in radiation domination, the Hubble rate is given by $H=T^2\sqrt{\pi^2 g_\ast /(90M_{\rm Pl}^2)}$ and includes the energy degrees of freedom, $g_\ast(T)$ and the reduced Planck mass, $M_{\rm Pl}$. The entropy of the SM is given by $s=2\pi^2g_{\ast s}T^3/45$ with $g_{\ast s}(T)$ the entropy degrees of freedom, while $T(a)$ is determined from entropy conservation, $a^3s(T)=\mathrm{const.}$; the QCD crossover is therefore included in the evolution. Both $g_\ast$ and $g_{\ast s}$ are tabulated and interpolated in $T$. We also normalize the scale factor by setting $a=1$ at the initial temperature $T_i=150\,\mathrm{GeV}$.

\section{A two-source freeze-in scenario}
\label{sec:benchmark}

We now introduce the particle physics scenario used in the numerical study. Its purpose is to provide a minimal setting in which two production sources, active at different times, populate distinct regions of the dark matter momentum distribution. We consider a real scalar dark matter particle $\chi$, odd under a $\mathbb Z_2$ symmetry, with all Standard Model fields being even. The dark matter is coupled feebly to the Standard Model through the Higgs portal and also receives a late contribution from the decay of a cold scalar condensate $\Phi$. The relevant terms in the potential are
\begin{equation}
\label{eq:benchmark_potential}
	V(H,\chi,\Phi)
	=
	\frac{1}{2}m_\chi^2\chi^2
	+
	\frac{\lambda}{4!}\chi^4
	+
	\frac{1}{2}m_\Phi^2\Phi^2
	+
	\frac{\lambda_{H\chi}}{2}H^\dagger H\chi^2
	+
	\frac{\mu_{\Phi\chi}}{2}\Phi\chi^2\,,
\end{equation}
where $H$ is the Higgs doublet of the SM, and it is given by $H=(0,h+v)^\intercal/\sqrt{2}$ in unitary gauge after the electroweak phase transition (EWPT). Here $h$ corresponds to the real mode, and $v$ to its VEV. $\lambda$ controls the elastic self-interaction, while $\lambda_{H\chi}$ controls the freeze-in production from the visible sector. After electroweak symmetry breaking, the Higgs portal interaction induces Higgs decays and annihilations to $\chi$ whenever kinematically allowed. We take the Higgs decay to be the dominant early-time contribution and neglect the production from annihilations, assuming $m_\chi\ll m_h$, where $m_h$ is the Higgs mass. 

The second source is described by the coupling $\mu_{\Phi\chi}\Phi\chi^2/2$, which allows the condensate component to decay as $\Phi\to\chi\chi$.\footnote{In principle, there is DM self-scattering mediated by the Higgs and the condensate. However, the decay rates are taken to be suppressed in the freeze-in limit, and these channels can be safely ignored.} We treat $\Phi$ as a cold nonthermal population with number density \(n_\Phi\) whose decay injects dark matter at a later time than the Higgs portal source, and we assume that its density is negligible such that it does not affect the expansion history of the Universe. Its phase-space distribution is taken to be concentrated at zero momentum,
\begin{equation}
        f_\Phi(t, k)
        =
        (2\pi)^3 n_\Phi(t)\delta^{(3)}( k).
\end{equation}
For the two-body decay, the injected \(\chi\) particles are monoenergetic, with $E_\chi^{(0)} = (m_\Phi^2+m_\chi^2-m_X^2)/(2m_\Phi)$. In our setup \(X=\chi\), so that \(\Phi\to\chi\chi\) and \(E_\chi^{(0)}=m_\Phi/2\). Assuming that \(\Gamma_\Phi\) denotes the decay rate of one condensate particle, the corresponding number and energy injection rates into the \(\chi\) sector are
\begin{equation}
        C_0^\Phi
        =
        2\,\Gamma_\Phi n_\Phi,
        \qquad
        C_E^\Phi
        =
        2E_\chi^{(0)}\Gamma_\Phi n_\Phi
        =
        m_\Phi\Gamma_\Phi n_\Phi.
\end{equation}
The decay width \(\Gamma_\Phi\) is treated as a free parameter chosen such that the injection occurs late relative to Higgs decay, while the initial value of \(n_\Phi\) is fixed so that this contribution to the final relic abundance remains subdominant. Treating the condensate as a cold and pressureless population with negligible backreaction on the expansion, its number density evolves as
\begin{equation}
    \dot n_\Phi +3Hn_\Phi = -\Gamma_\Phi n_\Phi\,.\label{eq:condensate_evolution}
\end{equation}
The solution determines the time dependence of \(n_\Phi(t)\) entering the condensate source terms in the cBE. This cold-condensate decay has been studied in DM production from inflaton or moduli-like condensates~\cite{Bhattacharya:2020zap,Ghosh:2022hen,Moroi:2020has,Ballesteros:2020adh}.

\begin{figure}[t!]
\centering
\includegraphics[width=0.81\textwidth]{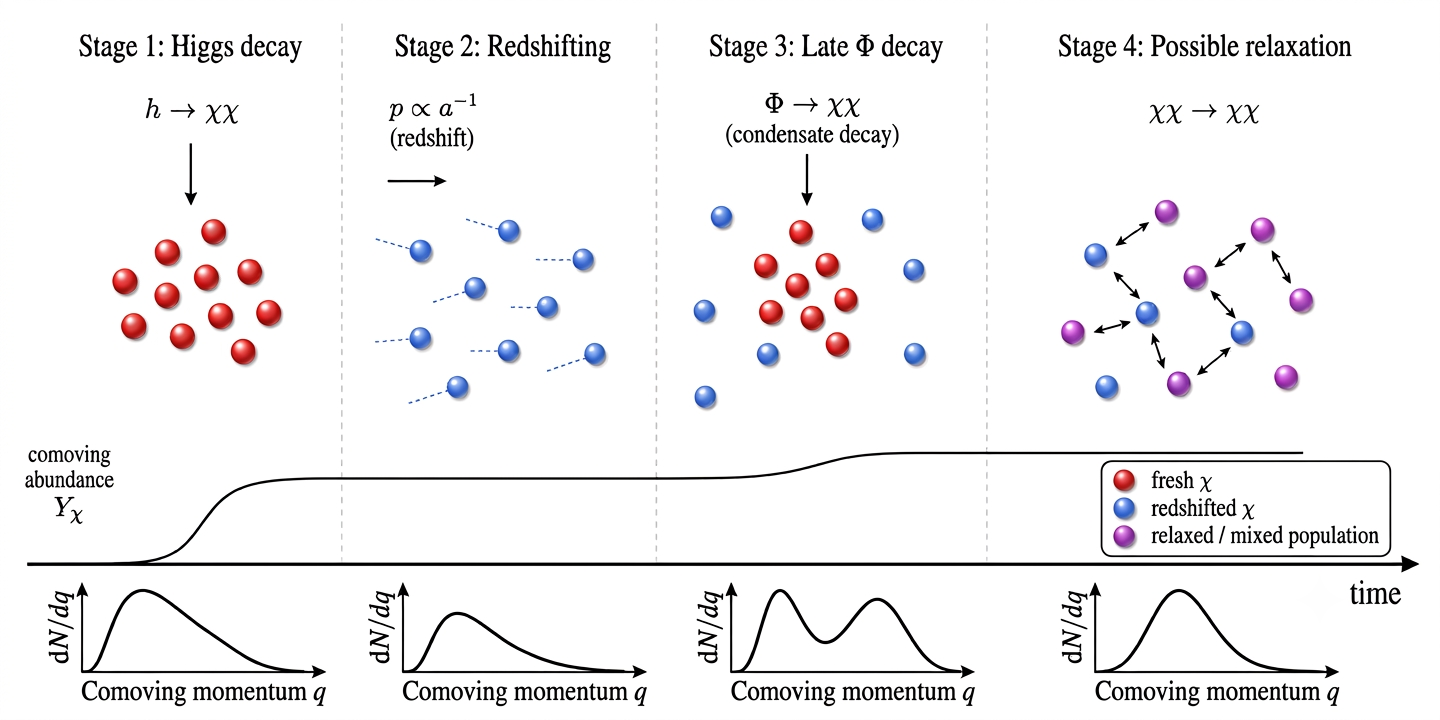}
\caption{Schematic illustration of the two-source freeze-in scenario studied in this work. Dark matter is first produced through Higgs decays from the thermal bath, after which the physical momenta redshift. At later times the cold condensate decays through \(\Phi\to\chi\chi\), injecting a second monoenergetic population with a different characteristic momentum. Elastic self-scattering may redistribute momentum between the two components and drive the distribution toward a thermal shape. The lower curve shows the qualitative evolution of \(Y_\chi\).
}
\label{fig:sketch}
\end{figure}

This scenario contains two freeze-in contributions,
\begin{equation}
\label{eq:CFI_two_sources}
	C_{\rm FI}
	=
	C_{h\to\chi\chi}
	+
	C_{\Phi\to\chi\chi} .
\end{equation}
The contribution of Higgs decay into DM to the collision operator is
\begin{equation}\label{eq:C_higgs_decay}
	 C_{h\to\chi\chi}(p)
	=
	\frac{(\lambda_{H\chi} v)^2}{8\pi g_\chi}\,\frac{T}{E p}\,\left(e^{-E_h^-(p)/T} -e^{-E_h^+(p)/T} \right)\,,
\end{equation}
where $E_h^\pm(p) = \left(m_h^2E \pm m_hp\sqrt{m_h^2 - 4m_\chi^2}\right)/(2m_\chi^2)$, $E = \sqrt{p^2+m_\chi^2}$, and $T$ is the temperature of the Standard Model bath. The second contribution comes from the decay of the cold condensate,
\begin{equation}
\label{eq:C_condensate_decay}
C_{\Phi\to\chi\chi}(p)
=
\frac{2\pi^2}{g_\chi}
\left(2\Gamma_\Phi n_\Phi\right)
\frac{\delta(p-p_0)}{p_0^2}\,,
\end{equation}
where $p_0= \frac{1}{2}\sqrt{m_\Phi^2-4m_\chi^2}$ is the physical momentum of each $\chi$ produced in the decay of a condensate particle at rest. In the absence of self-interactions, these two sources produce a bimodal distribution. The relative normalization of the two sources is fixed so that the total abundance reproduces the observed dark matter relic density, with the condensate contribution kept subdominant. This is illustrated in figure~\ref{fig:sketch}.

The contact interaction is generated by the quartic term. This isolates the effect of elastic self-scattering on the shape of the distribution. Other dark-sector models can lead to similar effective descriptions. For instance, scenarios in which dark-sector self-interactions can affect the production history include self-interacting sterile neutrinos produced via the freeze-in mechanism~\cite{Johns:2019cwc,Astros:2023xhe}, as well as axion-like particles~\cite{Badziak:2024qjg,Jain:2024dtw}. 

\section{The discretized Boltzmann equation}
\label{sec:numerics}

To study the relaxation of the freeze-in scenario introduced above, it is necessary to go beyond the fluid treatment and solve the fBE. To this end, it is convenient to introduce the comoving momentum $q \equiv a(t)p$, which removes the Hubble redshifting term from the left-hand side of the Boltzmann equation. We then evolve the distribution on a fixed grid in $q$, with the corresponding physical momenta given by $p_i=q_i/a(t)$. After discretizing the distribution in comoving-momentum bins, we write $f_i(t)\equiv f(t,q_i)$ and obtain the evolution equation from the fBE,
\begin{equation}
	\frac{d f_i}{dt}
	=
	C_{\text{el},i}[f] + C_{{\rm FI},i} .
\end{equation}

The collision operator $C_{\mathrm{FI},i}$ contains both the Higgs and condensate sources. For the condensate decay, at each time step the monoenergetic source at $q_0(a)=ap_0$ is split between the two neighboring grid points $q_L<q_0<q_R$ using the energy-linear deposition fractions $\alpha_R=(E_0-E_L)/(E_R-E_L)$ and $\alpha_L=1-\alpha_R$. The corresponding source amplitudes are normalized so that the discrete zeroth and energy moments reproduce exactly the injected particle-number and energy rates.
%
\subsection{The tensorial structure of the discretized operator}
\label{subsec:C}

The angular dependence in eq.~\eqref{eq:C22_general} can be reduced using the four-dimensional delta function, leaving an integral over the magnitudes of the momenta, together with a residual angular integral, which is included in the reduced kernel below. The operator has the gain-minus-loss structure
\begin{equation}
	C_\text{el}(p_i)
	=
	\frac{1}{2E_i g_\chi}
	\int dp_n\,dp_m\,
	\mathcal{F}(p_i,p_n,p_m)\,
	\Lambda_{inm},
\label{eq:reduced_collision_operator}
\end{equation}
where
\begin{equation}
\begin{aligned}
	\Lambda_{inm}
	={}
	f_n f_m
	\left(1+\eta_i f_i\right)
	\left(1+\eta_{\tilde j}f_{\tilde j}\right)-
	f_i f_{\tilde j}
	\left(1+\eta_n f_n\right)
	\left(1+\eta_m f_m\right),
\end{aligned}
\label{eq:reduced_statistical_factor}
\end{equation}
and $p_{\tilde j}$ is the fourth momentum fixed by energy conservation, 
\begin{equation}
	E_{\tilde j}=E_n+E_m-E_i,
	\qquad
	p_{\tilde j}^2=E_{\tilde j}^2-m_\chi^2 .
\label{eq:fourth_momentum}
\end{equation}
The function $\mathcal{F}$ denotes the reduced angular and kinematic kernel. It contains the squared matrix element, the angular phase-space measure, and the kinematic constraints.

We discretize the comoving momentum variable on a logarithmic grid, which is useful for distributions that span several momentum decades, as occurs when an early-produced population redshifts to low momenta before a later decay injects a distinct component,
\begin{equation}
	q_i \in [q_{\rm min},q_{\rm max}],
	\qquad i=1,\ldots,N_\text{grid} ,
\end{equation}
with quadrature weights $w_i$. For a fixed external momentum bin $i$, the reduced collision operator is evaluated by summing over two grid momenta, denoted by $n$ and $m$. The fourth momentum is then obtained from eq.~\eqref{eq:fourth_momentum}.

With this notation, the discretized operator reads
\begin{equation}
	C_{\text{el},i}[f]
	 = \sum_{n,m}
	W_{inm}
	\Lambda_{inm},
\label{eq:discrete_collision_operator}
\end{equation}
where
\begin{equation}
	W_{inm}
	=
	\frac{1}{2E_ig_\chi}\,
	\Delta p_n\,\Delta p_m\,
	\mathcal{F}_{inm} ,
\label{eq:discrete_weight}
\end{equation}
with $\Delta p_n$ and $\Delta p_m$ the physical momentum bin-widths. The explicit form of $\mathcal{F}_{inm}$ is given in appendix~\ref{app:collision_discretization}. The notation $\tilde j(i,n,m)$ emphasizes that the fourth momentum is fixed by energy conservation. We adopt the reduction in eq.~\eqref{eq:discrete_collision_operator} because the gain--loss structure maps naturally onto a single tensorized $\Lambda_{inm}$, which leads to a direct and convenient implementation with PyTorch. Its main drawback is that the fourth momentum $p_{\tilde j}$ is generally off-grid and therefore $f_{\tilde j}$ requires interpolation. Alternative parameterizations can avoid this interpolation and are likewise amenable to tensorization; these are discussed briefly in appendix~\ref{app:collision_discretization}.

The computational structure of the collision operator is explicit in eq.~\eqref{eq:discrete_collision_operator}. For each external momentum bin $i$, one has to sum over the two internal momentum indices $n$ and $m$, compute the corresponding fourth momentum, interpolate $f_{\tilde j}$ at that point, and evaluate the gain-minus-loss factor $\Lambda_{inm}$. In fact, the discretized operator has a simple computational scaling. For \(N_{\rm grid}\) momentum bins and \(N_\mu\) angular quadrature nodes used for the angular
integral in \(\mathcal{F}_{inm}\), one full evaluation of the raw operator requires \(\mathcal O(N_\mu N_{\rm grid}^3)\) arithmetic operations (or \(\mathcal O(N_{\rm grid}^3)\) in the case of an analytical kernel). The advantage of the tensorized formulation is that the \((i,n,m,\mu)\) contributions are independent before the final reductions and can be evaluated in parallel. The time integrator only receives the resulting vector $C_{\text{el},i}[f]$ at each step, so the cosmological evolution and other collision terms can be kept separate from the backend used to evaluate the self-collision operator.

Several checks are performed on the discretized operator. The zeroth (particle conservation) and first (energy conservation) moments of the elastic term must vanish,
\begin{equation}
	\sum_i w_i q_i^2 C_{\text{el},i}[f] \simeq 0 \qquad\text{and}\qquad \sum_i w_i q_i^2 E_i C_{\text{el},i}[f] \simeq 0\,,
\label{eq:number_conservation_discrete}
\end{equation}
up to numerical precision and boundary effects. These conditions are the discrete version of eq.~\eqref{eq:C22_conservation} and test that the numerical kernel redistributes particles in momentum space without changing the total number or energy density. Small violations can arise from the finite momentum range, interpolation near the grid boundaries, angular quadrature error, and finite precision. In the runs shown in section~\ref{sec:results}, these violations are controlled by projecting the elastic contribution onto the subspace with vanishing number and energy moments. Details of the projection are given in appendix~\ref{app:moments_conservation}. The projection preserves the conserved moments and the thermal fixed point of the elastic operator. Its size is quantified explicitly in figures~\ref{fig:app_projection_ngrid} and~\ref{fig:app_projection_pointwise} for a bimodal distribution given below and in figure~\ref{fig:app_projection_convergence_vs_x} throughout the evolution.

\subsection{Time-step integration}
\label{subsec:time_step_integration}

The time integration is performed using the scale factor as the independent variable, and since it spans many decades, we define $u \equiv \log a$, 
\begin{equation}
	\frac{d f_i}{du}
	=
	a\frac{d f_i}{da}
	=
	\frac{1}{H(a)}
	\left[
		C_{\text{el},i}[f]+C_{{\rm FI},i}
	\right] .
\label{eq:numerical_evolution_loga}
\end{equation}
The integration is split into two regimes. At early times, the self-scattering rate is negligible compared with the Hubble rate, and the evolution is driven only by the freeze-in source terms. In this regime, we set the self-collisions to zero, $C_{\text{el},i}[f]=0$, and evolve the distribution with a fourth-order Runge--Kutta method on fixed intervals in $u$ because the source terms are smooth. 

The code also monitors the effective self-scattering rate, $\Gamma_{2\to2}=n_\chi \langle\sigma_{\chi\chi\to\chi\chi}v\rangle$, and compares it with the Hubble rate. Once $\Gamma_{2\to2}/H > \epsilon_{\rm switch}=10^{-2}$, the elastic self-collision operator is included. At this stage the right-hand side becomes nonlinear in the distribution and can vary on the relaxation scale set by $\Gamma_{2\to2}^{-1}$. We therefore use an adaptive Heun method in $u=\log a$. The method first computes an Euler predictor and then corrects it using the average of the slopes at the beginning and end of the step. The difference between the predictor and corrected value provides a local error estimate used to adjust the step size. This gives a simple explicit integrator with two collision-operator evaluations per accepted step.

\section{Validation and performance}
\label{sec:performance}

For validation, we compare the elastic self-collision operator implemented in KineticXGPU with the corresponding semi-analytical evaluation provided by BEST~\cite{Yoon:2026rce}. Since the implementations use different radial discretizations and interpolation prescriptions, this comparison should be understood as a validation of the collision operator rather than as a direct comparison of complete Boltzmann solvers.

\begin{figure}[!t]
\centering
\includegraphics[width=0.66\textwidth]{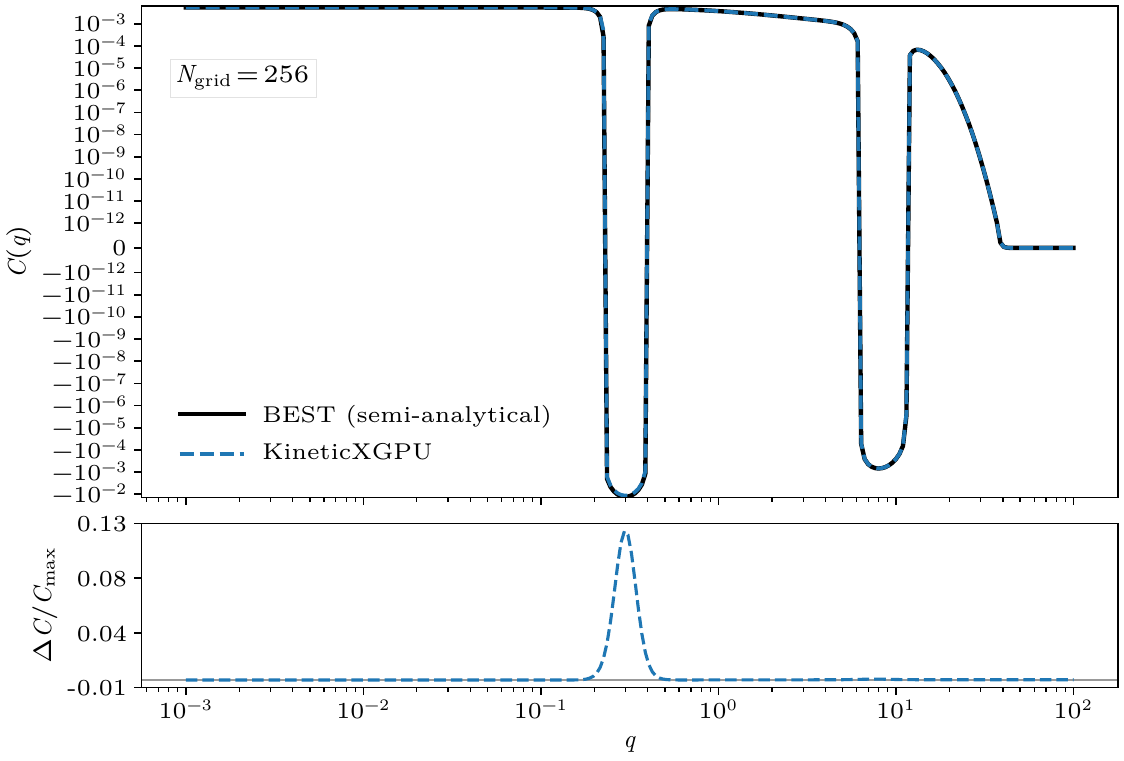}
\caption{Contact self-collision operator for the distribution in eq.~\eqref{eq:benchmark_two_bump} at $N_{\rm grid}=256$, with both collision operators evaluated in double precision. The lower panel shows the residual normalized to the maximum absolute BEST collision term. We use the analytical reduced kernel $\mathcal{F}^{\rm analytical}$ of eq.~\eqref{eq:app_F_contact_analytic}. The maximum absolute normalized residual decreases from $31.0\%$ at $N_{\rm grid}=96$ to $23.9\%$ at $N_{\rm grid}=128$ and $12.0\%$ at $N_{\rm grid}=256$. The projection discussed in appendix~\ref{app:moments_conservation} is not applied. The KineticXGPU result is multiplied by a factor of two to match the convention used by BEST.}
\label{fig:collision_validation}
\end{figure}

\begin{figure}[!t]
\centering
\includegraphics[width=0.72\textwidth]{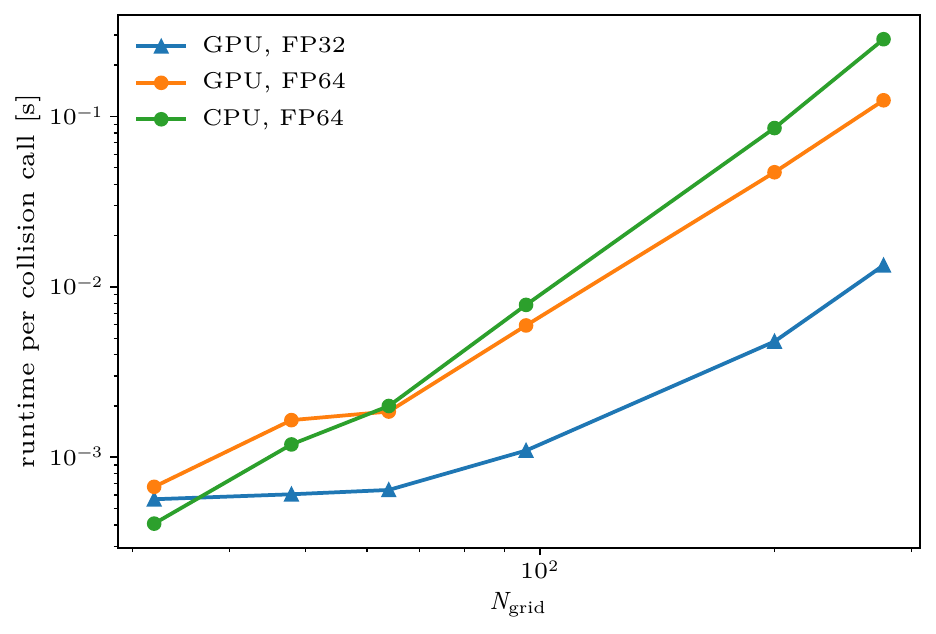}
\caption{Runtime per full-grid collision-operator call as a function of $N_{\rm grid}$ for CPU double precision and GPU single and double precision. All curves use the analytical reduced kernel $\mathcal{F}^{\rm analytical}$ of eq.~\eqref{eq:app_F_contact_analytic}, one batch with $B=N_{\rm grid}$, and no conservation projection. Each point is the median of 20 calls after 20 warm-up calls. For the GPU timings, CUDA is synchronized around each call and the compiled operator is warmed beforehand with inputs already resident on the device. Data transfers and compilation overhead are therefore excluded, while the analytical kernel is recomputed on every call and is not cached. The hardware is listed in table~\ref{tab:hardware}.
\label{fig:run_time}
}
\end{figure}

In figure~\ref{fig:collision_validation} we compare the collision operator obtained with KineticXGPU against the corresponding BEST evaluation for the following input distribution,
\begin{equation}
f(q)
=
\exp\left[
-\frac{\left(\ln(q/0.3)\right)^2}{2(0.18)^2}
\right]
+
0.5\,
\exp\left[
-\frac{\left(\ln(q/8)\right)^2}{2(0.22)^2}
\right] ,
\label{eq:benchmark_two_bump}
\end{equation}
which is a two-bump function. Additionally, we use
\begin{equation}\label{eq:bm_parameters}
        m_\chi = 1,
        \qquad
        \lambda = 1,\qquad
        a=1,
        \qquad
        q\in[10^{-3},10^2],
\end{equation}
so that the comoving and physical momenta coincide in the two solvers.  We use logarithmic spacing in the momentum grid. The two peaks are centered at $q/m_\chi=0.3$ and $q/m_\chi=8$. Since the first peak has occupation of order unity, the Bose-enhancement factors are numerically relevant and are included in this comparison. The largest residual occurs near the first nonthermal bump at \(q\simeq0.3\), where the collision integral is most sensitive to interpolation and radial grid discretization. The remaining difference is primarily a finite-grid discretization effect. KineticXGPU evaluates gain and loss on a common $(p_n,p_m)$ grid, whereas BEST discretizes the forward and backward contributions separately using different radial variables. Although equivalent in the continuum, these formulations sample the narrow low-momentum feature differently at finite resolution. We find that interpolation changes the result near this feature by about
$4.7\%$ at $N_{\rm grid}=96$, $2.7\%$ at $N_{\rm grid}=128$, and
$0.67\%$ at $N_{\rm grid}=256$, and is therefore subleading; the
momentum-domain boundary cut is negligible there. The decreasing residual with increasing $N_{\rm grid}$ reflects convergence toward the same continuum collision integral. An additional validation of the collision operator is provided in figure~\ref{fig:app_BE_fixed_point} with a Bose--Einstein distribution as input.

In KineticXGPU the discretized operator can be evaluated on either CPU or GPU. As an illustration, figure~\ref{fig:run_time} shows the runtime for one full-grid evaluation for the distribution in eq.~\eqref{eq:benchmark_two_bump} and parameters in eq.~\eqref{eq:bm_parameters}. The timings isolate a single evaluation of the elastic self-collision operator and therefore exclude source terms, adaptive step-size control, the conservation projection, and the cost of a complete cosmological evolution. This separates the performance of the tensorized collision kernel from model-dependent parts of the calculation, such as the freeze-in source, integration tolerances, and the number of accepted or rejected time steps. For context, BEST reports that a single Heun step for its semi-analytical $2\to2$ calculation with $N_{\rm grid}=272$ and with its integration parameter $N_{\rm eval}=10^6$ takes approximately $50\,{\rm seconds}$ using 272 MPI ranks~\cite{Yoon:2026rce}, while ref.~\cite{Hryczuk:2022gay} reports several hours of CPU time per parameter point for its phase-space calculation. These reports involve different algorithms, hardware, and computational tasks and are therefore not direct benchmarks against the single-call timings in figure~\ref{fig:run_time}; they illustrate the broader computational cost of phase-space Boltzmann calculations.



\section{Evolution}
\label{sec:results}

\begin{figure}[!tbp]
\centering
\includegraphics[width=\textwidth]{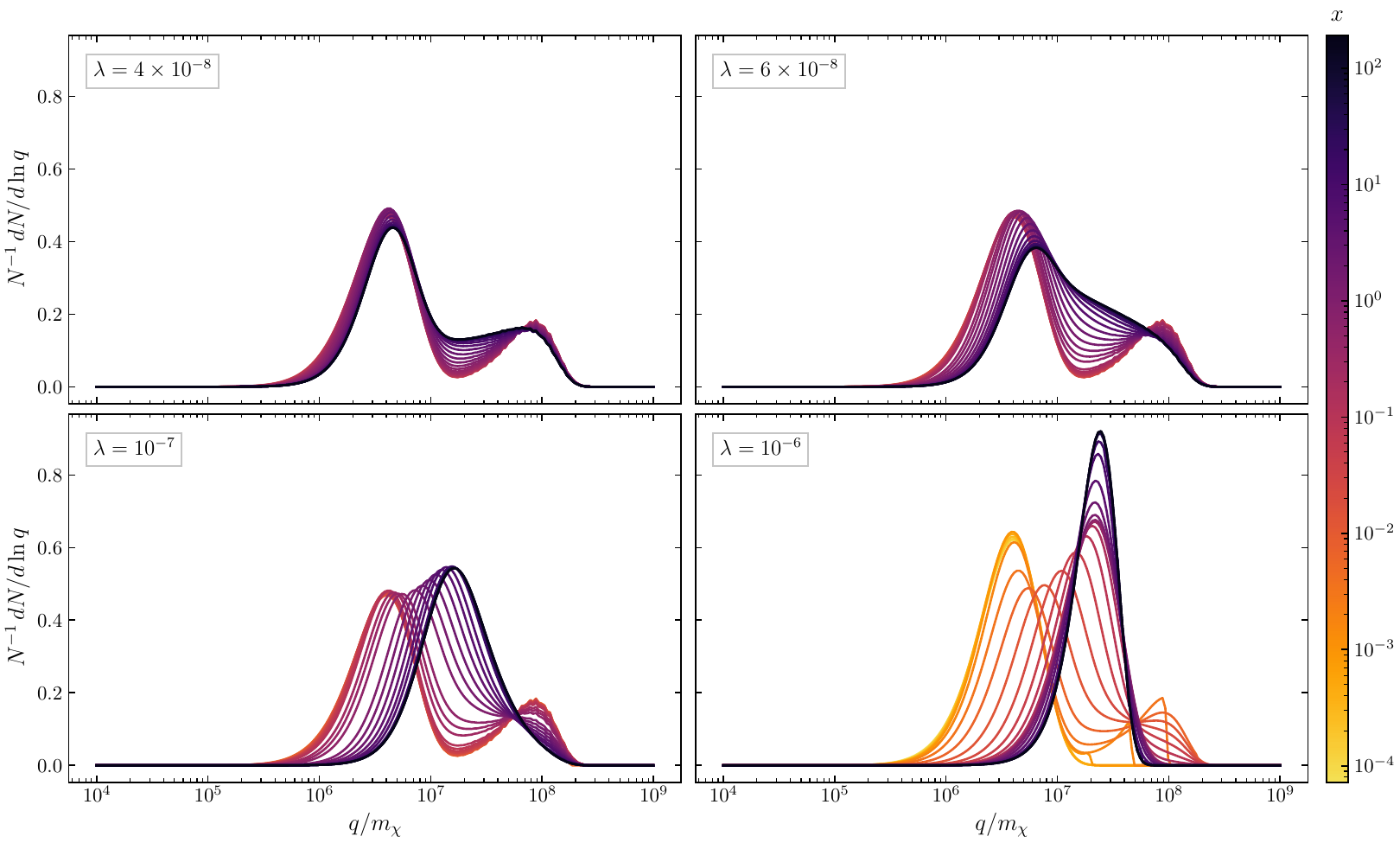}
\caption{Evolution of the normalized distribution, with $N=a^3n_\chi$. The color bar denotes time through $x=m_\chi/T$. All runs use $N_\text{grid}=200$, $q_\text{min}/m_\chi=10^4$, and $q_\text{max}/m_\chi=10^9$, with $\mathcal{F}^{\rm analytical}$ as defined in eq.~\eqref{eq:app_F_contact_analytic} and the conservation projection described in appendix~\ref{app:moments_conservation}. Dark matter is dilute for this benchmark, so quantum-statistical corrections to the collision operator are safely ignored. The total runtime of the evolution for $\lambda = 10^{-7}$ is 1.33 minutes, as documented in figure~\ref{fig:app_evolution_ngrid_convergence}. Larger couplings lead to larger runtimes because the stiffness requires smaller Heun steps.
}
\label{fig:evolution_examples}
\end{figure}

We now apply the numerical method to the two-source freeze-in scenario introduced in section~\ref{sec:benchmark}. The parameter set used in the runs is $m_\chi=100\,\text{keV}$, $\lambda_{H\chi}=7.84\times10^{-10}$, $m_\Phi=1.8\,\text{GeV}$, $\Gamma_\Phi=10^{-21}\,\text{GeV}$, and $n_{\Phi,0}=84.66\,\text{GeV}^3$, where $n_{\Phi,0}\equiv n_\Phi(a=1)$. These parameters reproduce the DM relic abundance, $\Omega_\chi h^2=0.12$~\cite{Planck:2018vyg}, with approximately $75\%$ and $25\%$ of the final number density produced by Higgs and condensate decays, respectively. The condensate energy density remains negligible and affects the cosmological history only by sourcing DM.\footnote{For this benchmark, we have checked that the backreaction on the expansion history is negligible. Evolving the condensate as a pressureless component according to eq.~\eqref{eq:condensate_evolution}, its energy density $\rho_\Phi=m_\Phi n_\Phi$ satisfies $\max(\rho_\Phi/\rho_{\rm rad})\sim\mathcal{O}(10^{-5})$, with the maximum reached around the decay epoch $H\sim\Gamma_\Phi$ (corresponding to $T\sim 50\,\text{MeV}$). The relativistic $\chi$ population produced by the condensate decay contributes only a few $\times10^{-5}$ of the radiation density, corresponding to an equivalent $\Delta N_{\rm eff}\sim\mathcal{O}(10^{-4})$ while the decay products are relativistic. The contribution of both the condensate and its decay products to the Hubble rate can therefore be safely neglected.} The evolution of the normalized distribution is shown in figure~\ref{fig:evolution_examples} for distinct values of the self-coupling. 

\begin{figure}[!tbp]
\centering
\includegraphics[width=0.71\textwidth]{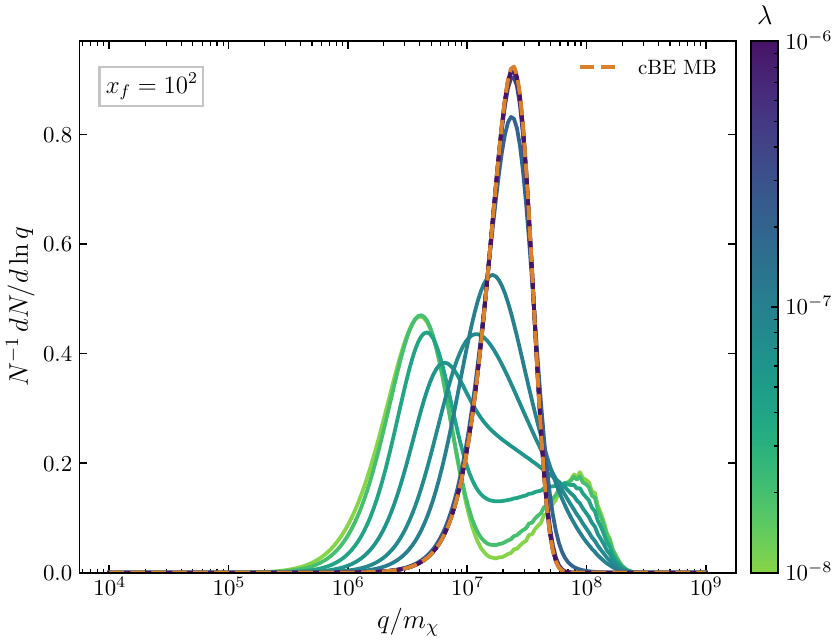}
\caption{Normalized phase-space distributions at $x_f=100$ for the self-coupling scan. The dashed curve shows the Maxwell--Boltzmann shape inferred from the cBE in eq.~\eqref{eq:cBE}. 
\label{fig:final_distributions}
}
\end{figure}

Defining $x=m_\chi/T$, the distributions at $x_f=100$ are shown in figure~\ref{fig:final_distributions}, where the dashed curve is the thermal distribution inferred from the cBE solution of eq.~\eqref{eq:cBE} at the same time. In the weakly self-interacting regime, the full phase-space solution remains visibly nonthermal, with a lower-momentum component from Higgs decays and a higher-momentum component from the later condensate decay.

\begin{figure}[!tbp]
\centering
\includegraphics[width=0.8\textwidth]{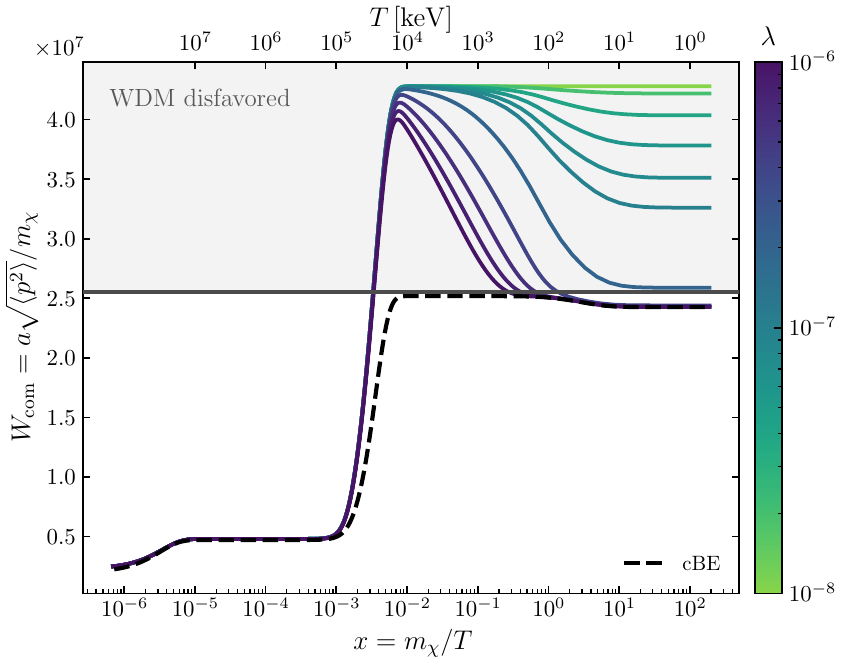}
\caption{Comoving warmness obtained from the full phase-space evolution and from the cBE. The factor $10^7$ applies only to the warmness values on the left vertical axis. The upper axis gives the Standard Model temperature. The solid horizontal line marks the WDM-equivalent reference associated with the $95\%$ C.L. limit $m_{\rm WDM}>5.7\,\text{keV}$ inferred from Lyman-$\alpha$ forest data at $z=4.2$--$5.0$~\cite{Irsic:2023equ}, and the lightly shaded region lies above that reference.
\label{fig:comoving_warmness}
}
\end{figure}

On the other hand, figure~\ref{fig:comoving_warmness} shows the comoving warmness $W_{\rm com}=a\sqrt{\langle p^2\rangle}/m_\chi=\sqrt{\langle q^2\rangle}/m_\chi$.\footnote{The average is defined as $\langle\mathcal O\rangle \equiv (g_\chi/n_\chi)\int_p\mathcal{O}f$.} Unlike a physical velocity evaluated at an arbitrary epoch, $W_{\rm com}$ approaches a constant once production and kinetic redistribution cease. In the weakly interacting solutions, the high-momentum component injected by condensate decay remains intact and contributes disproportionately to $\langle q^2\rangle$. The elastic collision term conserves particle number and energy, but not this second moment; i.e., when scattering is efficient, the energy carried by the narrow high-momentum component is redistributed over the full population, suppressing the tail and reducing the rms momentum. This explains why the collisionless solutions remain warmer, while the strongly interacting fBE solutions approach the thermal cBE result.

The two plateaus of the cBE curve, and of the fBE curves with sufficiently strong self-interactions, correspond to the relativistic and nonrelativistic regimes. Far from the transition, adiabatic evolution gives $T_\chi\propto a^{-1}$ in the relativistic regime and $T_\chi\propto a^{-2}$ in the nonrelativistic regime; in both limits $a\sqrt{\langle p^2\rangle}$ is therefore approximately constant. Around $x\sim1$, the dark matter becomes nonrelativistic and the relation between the conserved energy moment, pressure, and $\langle p^2\rangle$ changes, producing the drop between the two asymptotic plateaus. A collisionless distribution instead preserves each comoving momentum and therefore does not undergo the same kinetic redistribution. The horizontal line translates the $95\%$ C.L. lower bound $m_{\rm WDM}>5.7\,\text{keV}$ for a thermal relic~\cite{Irsic:2023equ} into the corresponding comoving warmness. This bound is inferred from Lyman-$\alpha$ forest spectra at $z=4.2$--$5.0$. Although the evolution shown ends at $x_f=100$ (corresponding to $T=1\,\mathrm{keV}$), the relevant comparison is with the post-production and post-relaxation plateau of $W_{\rm com}$, which remains constant once the collision terms become negligible. The shaded region is only a WDM-equivalent bound; it is not a model-independent Lyman-$\alpha$ exclusion on the values of $\lambda$. 

\begin{figure}[!htbp]
\centering
\includegraphics[width=0.75\textwidth]{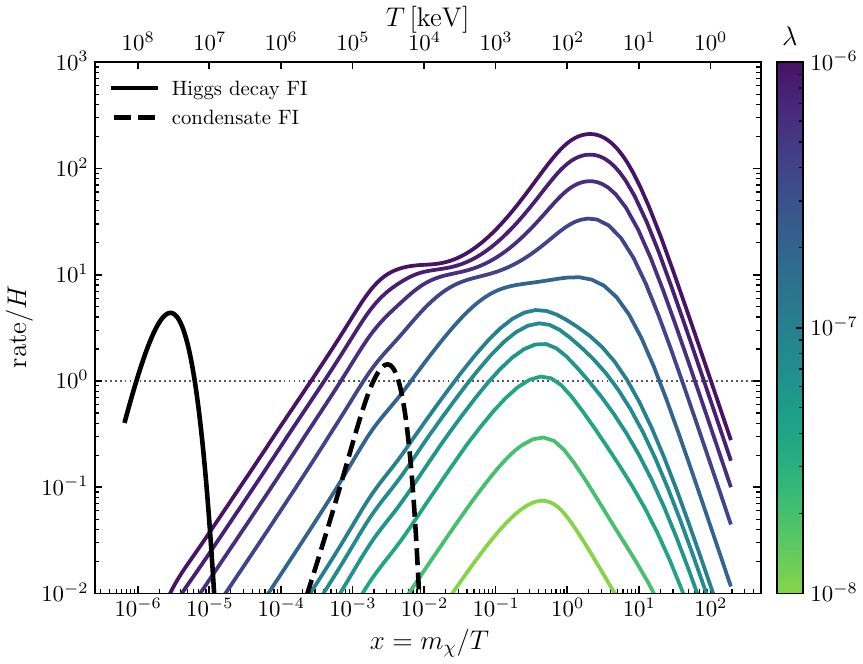}
\caption{Rates for the two-source scenario, normalized to the Hubble rate. The colored curves show the elastic self-scattering rate $\Gamma_{2\to2}=n_\chi\langle\sigma_{\chi\chi\to\chi\chi}v\rangle$.
\label{fig:interaction_rates}
}
\end{figure}

Figure~\ref{fig:interaction_rates} displays the $2\to2$ self-scattering rates normalized to the Hubble rate. It shows that the Higgs and condensate production rates are active in two separate windows. Increasing $\lambda$ raises the elastic rate and extends the interval over which $\Gamma_{2\to2}/H\gtrsim1$, explaining the progressive redistribution of the two momentum components in figures~\ref{fig:evolution_examples} and~\ref{fig:final_distributions}. The shallow dip at $x\sim10^{-2}$ ($T\sim50\,\mathrm{MeV}$) for the largest couplings coincides with condensate decay, which injects a narrow, high-momentum population. At the same time, the relativistic average $\langle\sigma_{\chi\chi\to\chi\chi}v\rangle$ is suppressed at larger pair energies. The resulting decrease in $\langle\sigma_{\chi\chi\to\chi\chi}v\rangle$ temporarily compensates the increase in $n_\chi$ supplied by the source, producing this feature in $\Gamma_{2\to2}/H$, which is most visible when self-scattering is already efficient and rapidly couples the injected population to the existing distribution; after injection and subsequent redshifting, the smooth evolution resumes its power-law scaling.

\clearpage
\section{Discussion and conclusions}
\label{sec:conclusions}

We have presented a tensorial implementation of the $2\to2$ self-collision operator entering the Boltzmann equation. The code is written in PyTorch and targets the tensor structure that appears after discretizing the collision integral, allowing the dominant operations to be evaluated efficiently in parallel. As demonstrated by the collision-operator performance results of section~\ref{sec:performance} and the evolution runtimes shown in figure~\ref{fig:app_evolution_ngrid_convergence}, this approach reduces the computational cost and hardware requirements to solve the phase-space Boltzmann equation with self-scatterings, making parametric scans feasible on a single GPU. Compared to existing phase-space approaches in which repeated collision-operator evaluations can require substantial CPU resources, the tensorial approach offers an optimized alternative. 

This tensorial treatment is useful when elastic self-scattering modifies the distribution but remains too weak to maintain local thermal equilibrium. In this regime, the phase-space distribution retains information about the production history, so observables relevant for structure formation cannot in general be inferred reliably from a thermal ansatz. Similar situations arise in a broad range of nonthermal dark-sector scenarios, including multi-source or sequential freeze-in, production from late decays, and self-interacting sterile-neutrino dark matter. ALPs may provide another example if elastic self-scatterings are present through a heavy mediator. Nonthermal populations produced during first-order phase transitions could also be studied once their subsequent evolution can be treated as homogeneous and isotropic. Different production mechanisms can be incorporated through the source term, while different elastic interactions enter through the corresponding collision kernel. The same numerical strategy can thus be applied to scalar, fermionic, and vector dark sectors whenever the phase-space distribution remains far from equilibrium. The resulting nonthermal distribution can then be used as input for structure-formation calculations of the matter power spectrum and Lyman-$\alpha$ observables. More generally, the tensorized strategy can be extended beyond elastic self-scattering to collision terms describing scattering with the Standard Model, annihilations, and other processes within the dark sector, after constructing the corresponding reduced kernels and gain--loss structures. Such extensions are currently being investigated for a future release.

As an application, we studied a two-source freeze-in scenario in which a scalar dark matter candidate is produced both through the Higgs portal and through the later decay of a cold condensate. In the absence of efficient self-interactions, the two sources populate different regions of momentum space, and the final distribution is not well described by a single thermal shape. Increasing the elastic self-coupling progressively smooths this structure and drives the solution toward the Maxwell--Boltzmann form, while reducing the comoving warmness as the distribution relaxes toward equilibrium. We showed this for a benchmark point in which sufficiently strong self-interactions reduce the late-time warmness, driving it below the WDM-equivalent reference associated with the Lyman-$\alpha$ bound. This illustrates the mechanism by which elastic self-interactions can weaken the free-streaming imprint and may alleviate structure-formation constraints on nonthermal frozen-in dark matter. Capturing this effect requires the full phase-space evolution whenever elastic self-scattering is too weak to maintain a thermal distribution. A systematic parameter scan, together with a calculation of the Lyman-$\alpha$ flux power spectra, would be required to translate the warmness into a quantitative constraint on this model.

\FloatBarrier

\acknowledgments
The author thanks Andrzej Hryczuk for useful discussions and comments on the manuscript. This work was supported by the National Science Centre (Poland) under the research Grant No. 2021/42/E/ST2/00009.



\appendix
\section{Comparison between the cBE and fBE output}
\label{app:solvers_comparison}

\begin{figure}[t]
\centering
\includegraphics[width=1\textwidth]{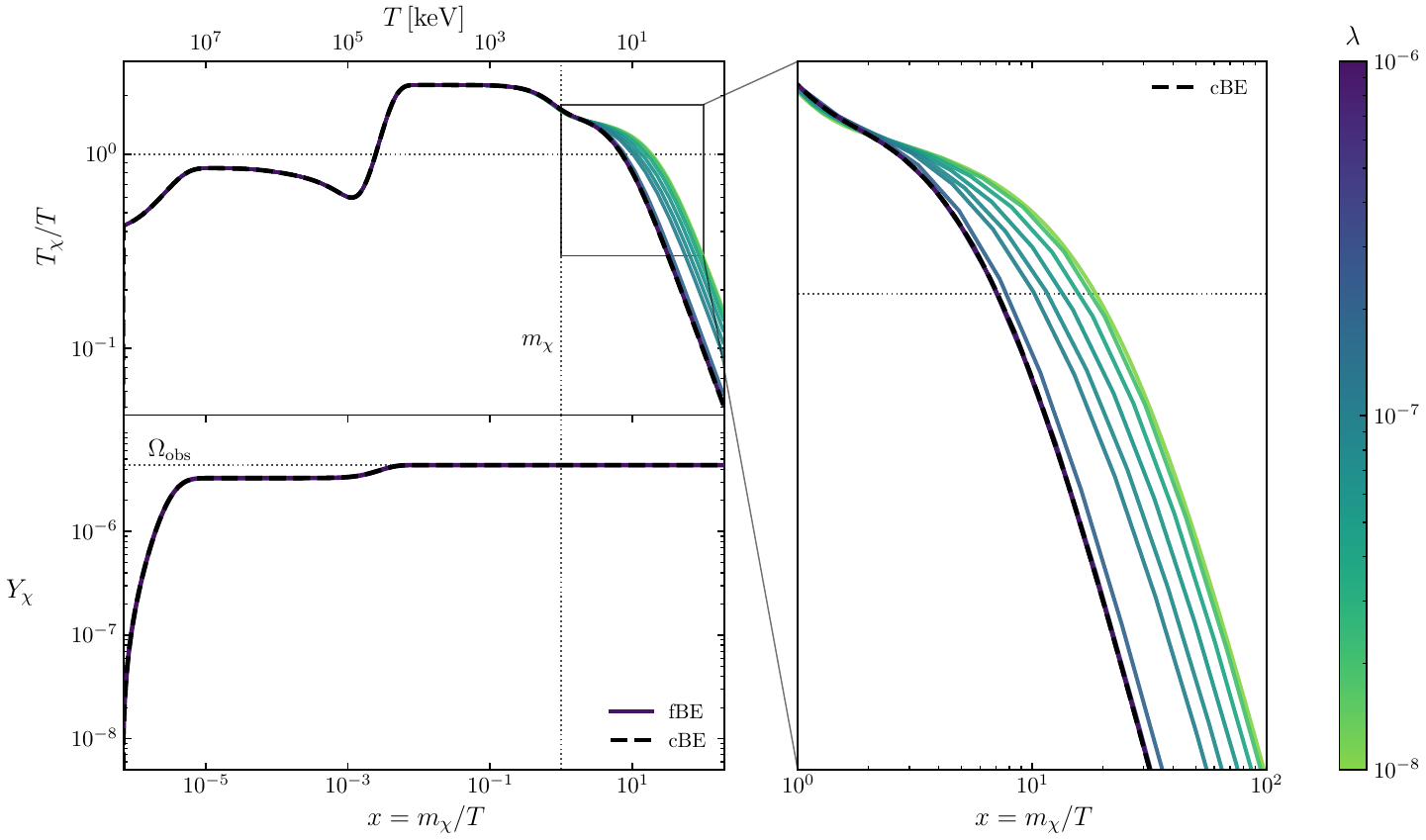}
\caption{Evolution of the yield (bottom) and temperature (top) obtained from the coupled Boltzmann equations in the fluid approximation, eq.~\eqref{eq:cBE} and the fBE, eq.~\eqref{eq:fbe}. The color bar denotes $\lambda$, with darker curves corresponding to larger couplings.
\label{fig:solution}
}
\end{figure}

The evolution of the yield and temperature obtained from the cBE and fBE is shown in figure~\ref{fig:solution}, while table~\ref{tab:fbe-cbe-moment-deviations} summarizes the final fBE temperature moment and comoving warmness. The temperature moment is defined as $T_\chi\equiv P_\chi/n_\chi=\langle p^2/(3E)\rangle$; for the Maxwell--Boltzmann ansatz this equals the temperature evolved by the cBE. We use $W_{\rm com}=a\sqrt{\langle p^2\rangle}/m_\chi$ because it approaches a constant once production and kinetic redistribution cease, making it a cleaner quantity for numerical comparisons and for connection with the WDM bound discussed in section~\ref{sec:results}.

\begin{table}[t]
\centering
\begin{tabular}{c c c c c c}
\hline
$\lambda$ & $\sigma_T/m_\chi$ [${\rm cm^2/g}$] & $T_\chi^{f,\,{\rm fBE}}$ [${\rm eV}$] & $W_{\rm com}^{f,\,{\rm fBE}}/10^7$ & $\Delta_T^f$ [\%] & $\Delta_W^f$ [\%] \\
\hline
$10^{-8}$ & $1.1\times10^{-10}$ & 293.83 & 4.2801 & 201.39 & 76.42 \\
$2\times10^{-8}$ & $4.3\times10^{-10}$ & 285.89 & 4.2199 & 193.25 & 73.95 \\
$4\times10^{-8}$ & $1.7\times10^{-9}$ & 262.24 & 4.0363 & 169.00 & 66.38 \\
$6\times10^{-8}$ & $3.9\times10^{-9}$ & 231.24 & 3.7829 & 137.19 & 55.93 \\
$8\times10^{-8}$ & $7.0\times10^{-9}$ & 200.15 & 3.5114 & 105.30 & 44.74 \\
$10^{-7}$ & $1.1\times10^{-8}$ & 173.42 & 3.2609 & 77.88 & 34.41 \\
$2\times10^{-7}$ & $4.3\times10^{-8}$ & 110.80 & 2.5885 & 13.66 & 6.70 \\
$4\times10^{-7}$ & $1.7\times10^{-7}$ & 98.63 & 2.4402 & 1.17 & 0.59 \\
$6\times10^{-7}$ & $3.9\times10^{-7}$ & 97.84 & 2.4304 & 0.35 & 0.18 \\
$8\times10^{-7}$ & $7.0\times10^{-7}$ & 97.65 & 2.4280 & 0.16 & 0.08 \\
$10^{-6}$ & $1.1\times10^{-6}$ & 97.58 & 2.4271 & 0.09 & 0.04 \\
\hline
\end{tabular}
\caption{Final-state fBE temperature moment and comoving warmness for the 100-keV self-coupling scan, at $x_f=100$ and $N_{\rm grid}=200$. For each nonthermal fBE distribution, $T_\chi=P_\chi/n_\chi=\langle p^2/(3E)\rangle$ and $W_{\rm com}=a\sqrt{\langle p^2\rangle}/m_\chi=\sqrt{\langle q^2\rangle}/m_\chi$. The cBE reference uses its evolved $T_\chi^f=97.49\,{\rm eV}$ and has $W_{\rm com,cBE}^f/10^7=2.4260$. We define $\Delta_T^f=|T_\chi^{\rm fBE}/T_\chi^{\rm cBE}-1|$ and $\Delta_W^f=|W_{\rm com}^{\rm fBE}/W_{\rm com}^{\rm cBE}-1|$; both deviations are percentages. The transfer cross section uses $\sigma_T/m_\chi=\lambda^2/(64\pi m_\chi^3)$ with $m_\chi=100\,{\rm keV}$.}
\label{tab:fbe-cbe-moment-deviations}
\end{table}

\section{Discretization of the self-collision operator}
\label{app:collision_discretization}

Here we describe the discretization of the elastic self-collision operator used in the numerical solution of the phase-space Boltzmann equation. The construction follows the reduction of the $2\to2$ self-scattering operator described in ref.~\cite{Hryczuk:2022gay} with two differences. First, we use the relativistic energy-conservation condition to determine the fourth momentum. Second, for the contact interaction, the remaining angular kernel can be evaluated in closed form. Note that alternative parameterizations of full \(2\to2\) collision terms are possible. For instance, ref.~\cite{Ala-Mattinen:2022nuj} uses a forward/backward decomposition that avoids interpolating the unknown distribution, while ref.~\cite{Aboubrahim:2023yag} exploits an analytic reduction of the elastic kernel for matrix elements depending on one Mandelstam variable. These formulations can also be cast in tensorial form, although with the additional kinematic bookkeeping associated with their respective parameterizations. For the identical-particle self-scattering, the reduction below gives a kernel that is further simplified for a contact interaction, but the same tensorial strategy can be applied to more general kernels.

We work with the comoving momentum variable, $q = a(t)p$, and discretize it on a logarithmic grid $q_i$ with $i=1,\ldots,N_{\rm grid}$. The corresponding physical momenta and energies are $p_i = q_i/a$ and $E_i = \sqrt{p_i^2+m_\chi^2}$, while the physical momentum width associated with the grid point $i$ is denoted by $\Delta p_i = \Delta q_i/a.$
It is useful to define the moment weight
\begin{equation}
        h_i \equiv p_i^2 \Delta p_i \,,
\end{equation}
so that the discrete number and energy moments of a
distribution are proportional to
\begin{equation}
        \sum_i h_i f_i ,
        \qquad
        \sum_i h_i E_i f_i ,
\end{equation}
respectively, up to the common factor $g_\chi/(2\pi^2)$.

For a fixed external momentum $p_i$, the reduced elastic collision operator can be written as in eq.~\eqref{eq:reduced_collision_operator}. The kernel ${\cal F}$ contains the reduced angular phase-space measure, the squared matrix element, and the kinematic constraints. The fourth momentum is fixed by energy conservation, $E_{\tilde j}=E_n + E_m - E_i$ and $p_{\tilde j}^2=E_{\tilde j}^2 - m_\chi^2 $. Only configurations satisfying $E_{\tilde j}\geq m_\chi$ and $q_{\rm min}\leq q_{\tilde j}\leq q_{\rm max}$ are kept in the numerical sum. After using the spatial delta function, the remaining radial measures are $p_n^2dp_n/(2E_n)$ and $p_m^2dp_m/(2E_m)$. Integrating the energy delta function over the relative azimuth gives $|\partial(E_i+E_{\tilde j}-E_n-E_m)/\partial\phi| =p_np_m\sqrt{1-\mu^2}\sqrt{1-\nu^2}|\sin\phi|/E_{\tilde j}$. The two azimuthal roots cancel the factor $1/(2E_{\tilde j})$, leaving the factor $p_np_m/(E_nE_m)$ in the reduced kernel. The reduced operator can then be written as
\begin{equation}
        {\cal F}(p_i,p_n,p_m)
        =
        \frac{1}{4(2\pi)^4}\frac{p_np_m}{E_nE_m}
        \int_{-1}^{1} d\mu
        \int_{-1}^{1} d\nu\,
        \frac{|\mathcal{M}|^2}
        {\sqrt{1-\mu^2}\sqrt{1-\nu^2}\sqrt{1-\cos^2\phi}} ,
\label{eq:app_F_2d}
\end{equation}
where $\mu \equiv \cos\theta_2$ and $\nu \equiv \cos\theta_3$.
Here $\theta_2$ is the angle between $p_i$ and
$p_n$, while $\theta_3$ is the angle between $p_i$ and
$p_m$. The angle $\phi$ is the relative azimuthal angle between
the projections of $p_n$ and $p_m$ onto the plane
orthogonal to $p_i$. Energy conservation fixes
\begin{equation}
        \cos\phi
        =
        \frac{
        m_\chi^2
        +E_nE_m-E_nE_i-E_mE_i
        +p_i(p_n\mu+p_m\nu)
        -p_np_m\mu\nu
        }
        {p_np_m\sqrt{1-\mu^2}\sqrt{1-\nu^2}} .
\label{eq:app_cosphi}
\end{equation}
The physical integration region is restricted by $|\cos\phi|\leq 1$. For the contact interaction, $|\mathcal{M}|^2$ is momentum independent. In this case, the integral over $\nu$ can be performed analytically. For fixed $\mu$, define
\begin{align}
        A_0
        &=
        m_\chi^2
        +E_nE_m-E_nE_i-E_mE_i
        +p_i p_n \mu ,
        \\
        A_1
        &=
        p_i p_m - p_n p_m\mu ,
        \\
        K^2
        &=
        p_n^2 p_m^2(1-\mu^2) .
\end{align}
Then $1-\cos^2\phi \geq 0$ is equivalent to $a_\nu\nu^2+b_\nu\nu+c_\nu \geq 0$ with
\begin{equation}
        a_\nu = -(K^2+A_1^2),
        \qquad
        b_\nu = -2A_0A_1,
        \qquad
        c_\nu = K^2-A_0^2 .
\end{equation}
Using eq.~\eqref{eq:app_cosphi}, the angular measure can be rewritten as
\begin{equation}
        \frac{d\mu\,d\nu}
        {\sqrt{1-\mu^2}\sqrt{1-\nu^2}\sqrt{1-\cos^2\phi}}
        =
        p_n p_m
        \frac{d\mu\,d\nu}
        {\sqrt{a_\nu\nu^2+b_\nu\nu+c_\nu}} .
\label{eq:app_measure_rewrite}
\end{equation}

To further reduce the integral, let $\Delta_\nu = b_\nu^2 - 4a_\nu c_\nu $. If $\Delta_\nu<0$, or if the allowed interval does not overlap with
$-1\leq\nu\leq 1$, the contribution is set to zero. Otherwise, the roots
\begin{equation}
        r_\pm =
        \frac{-b_\nu\pm\sqrt{\Delta_\nu}}{2a_\nu}
\end{equation}
are clipped to the physical interval $[-1,1]$, giving the integration
limits $\nu_-$ and $\nu_+$. The analytic integral over $\nu$ is then
\begin{equation}
        {\cal I}(\mu)
        =
        \frac{1}{\sqrt{-a_\nu}}
        \left[
        \sin^{-1}
        \left(
        \frac{2a_\nu\nu_-+b_\nu}{\sqrt{\Delta_\nu}}
        \right)
        -
        \sin^{-1}
        \left(
        \frac{2a_\nu\nu_++b_\nu}{\sqrt{\Delta_\nu}}
        \right)
        \right] .
\label{eq:app_I_mu}
\end{equation}
The reduced kernel is 
\begin{equation}
        {\cal F}_{inm}
        =
        \frac{|\mathcal{M}|^2}{4(2\pi)^4}\frac{(p_np_m)^2}{E_nE_m}
        \int_{-1}^{1} d\mu\,{\cal I}(\mu) .
\label{eq:app_F_1d}
\end{equation}
For a constant matrix element, the remaining angular kernel can be reduced to a closed-form interval-overlap expression. To see this, define $Q(\mu)\equiv\sqrt{p_i^2+p_n^2-2p_i p_n\mu} $. Using energy conservation, one finds $K^2 + A_1^2 = p_m^2 Q^2(\mu)$, so that, whenever the physical interval in $\nu$ is non-empty, the integral in eq.~\eqref{eq:app_I_mu} is equivalently
\begin{equation}
        {\cal I}(\mu)
        =
        \frac{\pi}{p_m Q(\mu)} .
\end{equation}
The remaining condition is purely kinematic. The allowed range of $Q$ is the overlap of the two intervals
\begin{equation}
        |p_i-p_n| \leq Q \leq p_i+p_n ,
        \qquad
        |p_m-p_{\tilde j}| \leq Q \leq p_m+p_{\tilde j} .
\end{equation}
Equivalently, defining
\begin{align}
        Q_- &=
        \max\left(
        |p_i-p_n|,\,
        |p_m-p_{\tilde j}|
        \right) ,
        \\
        Q_+ &=
        \min\left(
        p_i+p_n,\,
        p_m+p_{\tilde j}
        \right) ,
\end{align}
the contact-interaction kernel can be written in closed form as
\begin{equation}
        {\cal F}_{inm}^{\rm analytical}
        =
        \frac{|\mathcal{M}|^2}{64\pi^3 p_i}\frac{p_np_m}{E_nE_m}
        \left[ Q_+ - Q_- \right]_+ ,
\label{eq:app_F_contact_analytic}
\end{equation}
where $[x]_+ \equiv \max(x,0)$ and $p_{\tilde j}$ is determined by energy conservation. This expression provides a useful validation of the angular integration and can be used as an optimized kernel for the contact case. 

Using the weight defined in eq.~\eqref{eq:discrete_weight}, the discrete collision operator reads
\begin{equation}
        C_{{\rm el},i}^{\rm raw}
        =
        \sum_{n,m}
        W_{inm}\Lambda_{inm} .
\label{eq:app_dense_collision}
\end{equation}
Since $q_{\tilde j}$ does not generally coincide with a grid point, the value of the distribution at this momentum has to be interpolated. If $q_{j_L}\leq q_{\tilde j}\leq q_{j_R}$, we define energy-linear weights
\begin{equation}
        \omega_R
        =
        \frac{E_{\tilde j}-E_{j_L}}
        {E_{j_R}-E_{j_L}},
        \qquad
        \omega_L = 1-\omega_R .
\label{eq:app_interp_weights}
\end{equation}
For the full quantum-statistical operator, we define
\begin{equation}
y_k \equiv \log\left(\frac{f_k}{1+\eta f_k}\right),
\qquad
y_{\tilde j}
=
\omega_L y_{j_L}+\omega_R y_{j_R},
\qquad
f_{\tilde j}
=
\frac{e^{y_{\tilde j}}}{1-\eta e^{y_{\tilde j}}}.
\label{eq:app_log_interp}
\end{equation}
This choice preserves the thermal form $\log [f/(1+\eta f)]={\rm const}-E/T$ under interpolation and improves the cancellation of the gain and loss terms near LTE (cf. figure~\ref{fig:app_BE_fixed_point}). 

\section{Moment conservation of the $2\to2$ operator}
\label{app:moments_conservation}

The continuum elastic $2\to2$ collision operator conserves the particle number and energy stored in the dark sector. In the discretized form used in eq.~\eqref{eq:app_dense_collision}, these conservation laws are not satisfied exactly on a finite momentum grid. The residual violations arise from the finite momentum range, the interpolation of the off-grid momentum fixed by energy conservation, the numerical angular quadrature, and floating-point precision. We therefore use a two-moment projection of the elastic operator. Similar conservation-enforcing strategies are common in plasma simulations~\cite{2015JCoPh.297..357T}. 

For a discrete collision operator $C_i$, we define the normalized number and energy moment residuals
\begin{equation}
\epsilon_N[C]
=
\frac{\sum_i h_i C_i}
{\sum_i h_i |C_i|},
\qquad
\epsilon_E[C]
=
\frac{\sum_i h_i E_i C_i}
{\sum_i h_i E_i |C_i|}.
\label{eq:app_epsilon_definition}
\end{equation}
The quantities $\epsilon_N^{\rm raw}$ and $\epsilon_E^{\rm raw}$ denote eq.~\eqref{eq:app_epsilon_definition} evaluated on the raw operator, whereas $\epsilon_N^{\rm proj}$ and $\epsilon_E^{\rm proj}$ denote the same quantities evaluated after projection. For the raw operator we define
\begin{equation}
R_0 =
\sum_i h_i C_{{\rm el},i}^{\rm raw},
\qquad
R_1 =
\sum_i h_i E_i C_{{\rm el},i}^{\rm raw}.
\end{equation}
The projected operator is taken to be
\begin{equation}
C_{{\rm el},i}^{\rm proj}
\equiv
C_{{\rm el},i}^{\rm raw}
-
\xi_i(\alpha+\beta E_i) ,
\label{eq:app_self_projection}
\end{equation}
where $\xi_i$ restricts the correction to the support of the distribution.
In the numerical runs we use
\begin{equation}
\xi_i =
\begin{cases}
|f_i|/f_{\rm max},
&
|f_i|/f_{\rm max}>\epsilon_{\rm supp},\\
0,
&
|f_i|/f_{\rm max}\leq\epsilon_{\rm supp}.
\end{cases}
\end{equation}
Here $f_{\rm max}\equiv \text{max}_i \vert f_i\vert$, while $\epsilon_{\rm supp}$ is a numerical support cutoff used to restrict the conservation projection to momentum bins where the distribution is appreciably populated. It prevents the projection correction from acting on empty bins.

The constants $\alpha$ and $\beta$ are fixed by imposing
\begin{equation}
\sum_i h_i C_{{\rm el},i}^{\rm proj}=0,
\qquad
\sum_i h_i E_i C_{{\rm el},i}^{\rm proj}=0 .
\end{equation}
Equivalently, defining
\begin{equation}
A_{00}=\sum_i h_i\xi_i,
\qquad
A_{01}=\sum_i h_iE_i\xi_i,
\qquad
A_{11}=\sum_i h_iE_i^2\xi_i ,
\end{equation}
one solves
\begin{equation}
\begin{pmatrix}
A_{00} & A_{01} \\
A_{01} & A_{11}
\end{pmatrix}
\begin{pmatrix}
\alpha\\
\beta
\end{pmatrix}
=
\begin{pmatrix}
R_0\\
R_1
\end{pmatrix}.
\label{eq:app_projection_system}
\end{equation}
Explicitly,
\begin{equation}
\alpha =
\frac{R_0A_{11}-R_1A_{01}}
{A_{00}A_{11}-A_{01}^2},
\qquad
\beta =
\frac{R_1A_{00}-R_0A_{01}}
{A_{00}A_{11}-A_{01}^2}.
\end{equation}
\begin{figure}[t!]
\centering
\includegraphics[width=0.95\textwidth]{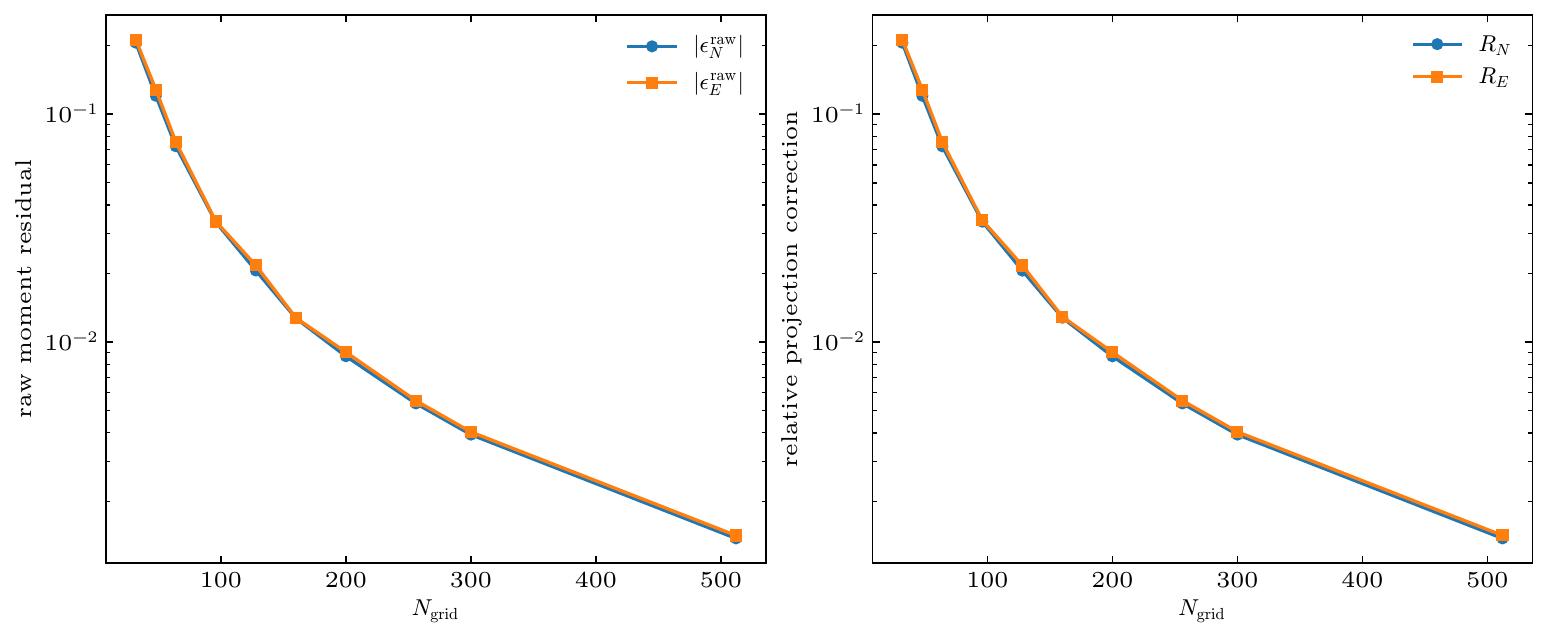}
\caption{Projection diagnostics for the same two-bump function defined in eq.~\eqref{eq:benchmark_two_bump} of section~\ref{sec:performance}. The left panel shows the absolute raw number and energy moment residuals defined in eq.~\eqref{eq:app_epsilon_definition} as a function of the number of momentum grid points. The right panel shows the corresponding relative projection corrections $R_N$ and $R_E$, defined in eq.~\eqref{eq:app_projection_relative_correction}. Both the raw moment residuals and the projection corrections decrease systematically with increasing $N_{\rm grid}$, showing convergence toward the continuum conservation laws and that the projection becomes progressively negligible at higher resolution.
}
\label{fig:app_projection_ngrid}
\end{figure}
\begin{figure}[t!]
\centering
\begin{minipage}{0.49\textwidth}
    \centering
    \includegraphics[width=\textwidth]{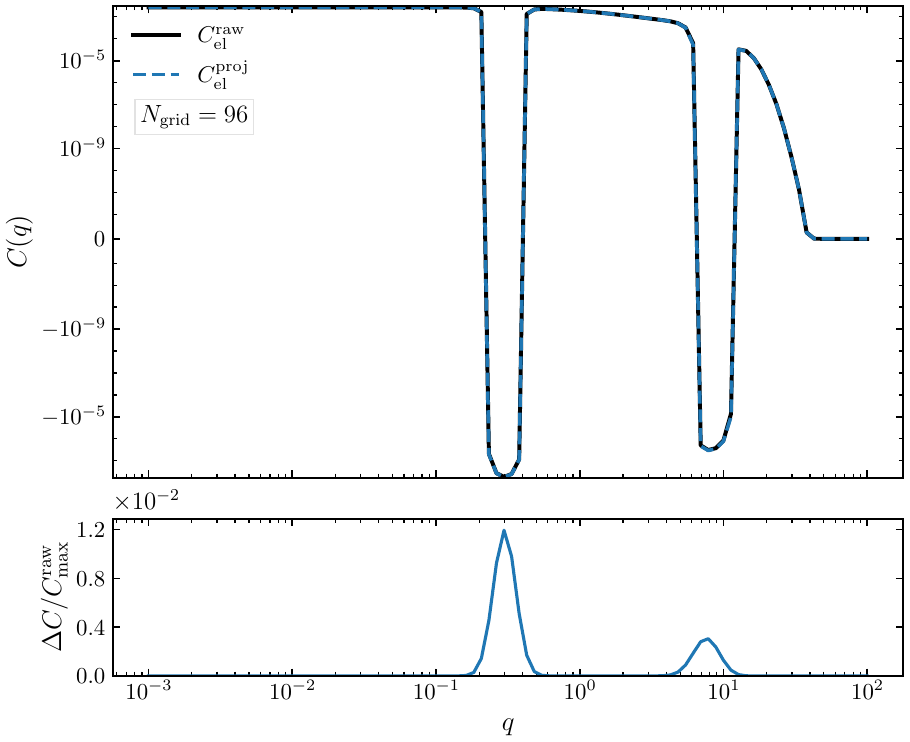}
\end{minipage}
\hfill
\begin{minipage}{0.49\textwidth}
    \centering
    \includegraphics[width=\textwidth]{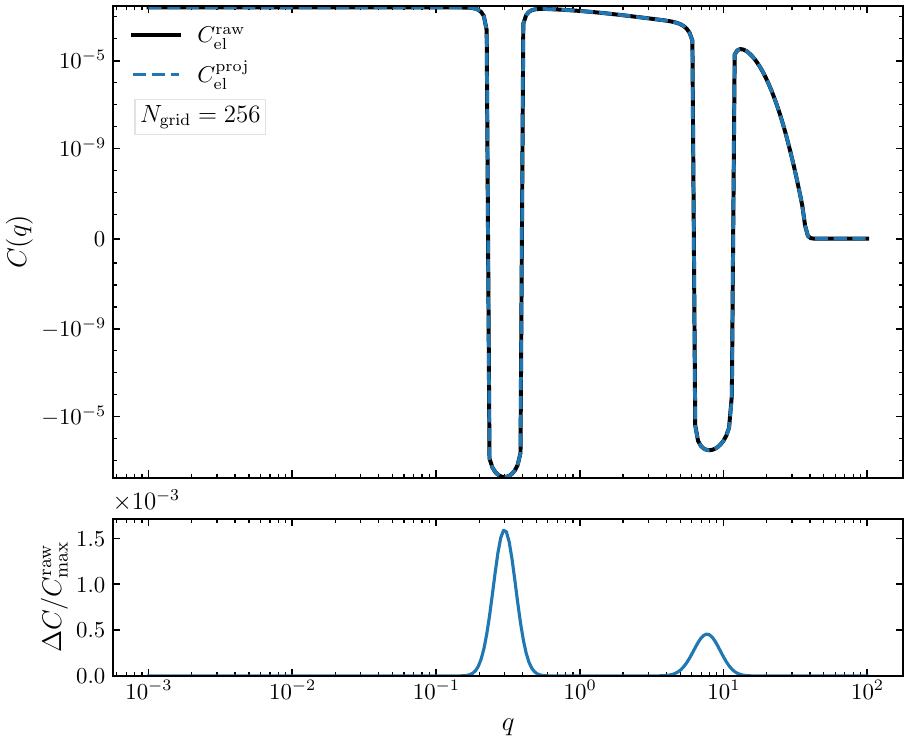}
\end{minipage}
\caption{Pointwise comparison of the raw and projected collision operators for the same two-bump distribution used in the independent-implementation comparison. The left panel uses \(N_{\rm grid}=96\), while the right panel uses \(N_{\rm grid}=256\). Both are evaluated with the analytical reduced kernel \(\mathcal{F}^{\rm analytical}\) of eq.~\eqref{eq:app_F_contact_analytic}. The upper panels show \(C_{\rm el}^{\rm raw}(q)\) and \(C_{\rm el}^{\rm proj}(q)\), while the lower panels show \(\Delta C/C_{\rm max}^{\rm raw}\), with \(\Delta C=|C_{\rm el}^{\rm proj}-C_{\rm el}^{\rm raw}|\) and \(C_{\rm max}^{\rm raw}=\max_i|C_{{\rm el},i}^{\rm raw}|\). The projected operator enforces the conservation of number and energy moments. The pointwise correction remains small relative to the peak amplitude of the raw collision operator and decreases as the momentum-grid resolution is increased.
}
\label{fig:app_projection_pointwise}
\end{figure}
A clarification regarding the projection is useful here. This projection does not represent an additional physical interaction. It removes only the component of the discretized elastic operator that would otherwise change the two conserved moments. Also, note that the size of the raw residuals is not a fractional error on the relic abundance or on the final distribution. The quantities $\epsilon_N^{\rm raw}$ and $\epsilon_E^{\rm raw}$ measure the degree to which the instantaneous raw operator fails to cancel its gain and loss contributions after summing over the finite grid. The projection does not impose a thermal fixed point but acts only on the discrete collision operator by subtracting the two components proportional to \(\xi_i\) and \(\xi_i E_i\) required to enforce the number and energy moments; i.e., no Maxwell--Boltzmann ansatz is used, and no temperature or chemical potential is fitted. That is, the projection preserves the two conserved moments of the discretized operator; it does not thermalize the distribution pointwise or erase nonthermal structure by construction. Its possible numerical effect is limited to the projected relaxation. Therefore, precision applications to Lyman-\(\alpha\) constraints or other observables would require additional validation of the final nonthermal shape, for example, with larger momentum grids or a more explicitly conservative discretization.

\clearpage

To quantify the size of the projection correction itself, we define the number- and energy-weighted relative corrections
\begin{equation}
R_N
=
\frac{
\sum_i h_i
\left|
C_{{\rm el},i}^{\rm proj}
-
C_{{\rm el},i}^{\rm raw}
\right|
}{
\sum_i h_i
\left|
C_{{\rm el},i}^{\rm raw}
\right|
},
\qquad
R_E
=
\frac{
\sum_i h_i E_i
\left|
C_{{\rm el},i}^{\rm proj}
-
C_{{\rm el},i}^{\rm raw}
\right|
}{
\sum_i h_i E_i
\left|
C_{{\rm el},i}^{\rm raw}
\right|
}.
\label{eq:app_projection_relative_correction}
\end{equation}
These quantities measure how intrusive the projection is in the same number- and energy-weighted norms used to test moment conservation. Figure~\ref{fig:app_projection_ngrid} shows the dependence of the raw moment residuals and the projection correction on the momentum-grid resolution for the two-bump function in eq.~\eqref{eq:benchmark_two_bump}. Both quantities decrease systematically with increasing $N_{\rm grid}$. On the other hand, figure~\ref{fig:app_projection_pointwise} shows that the projection does not reshape the collision operator. When normalized to the peak value of the raw operator, the maximal pointwise correction is $1.19\%$ at $N_{\rm grid}=96$ and $0.16\%$ at $N_{\rm grid}=256$, demonstrating its decrease with grid refinement.

\begin{figure}[t!]
    \centering
    \includegraphics[width=0.7\textwidth]
    {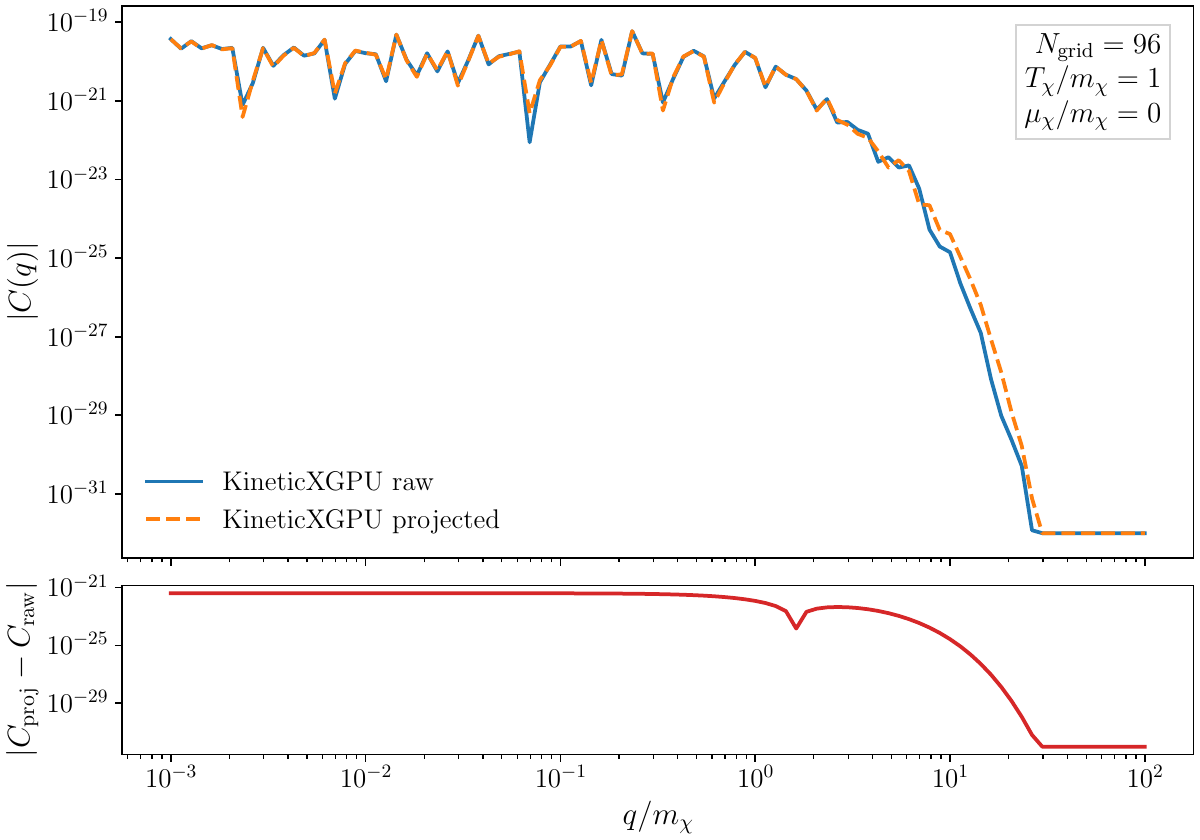}
    \caption{
Bose--Einstein fixed-point test of the elastic self-collision operator for $N_{\rm grid}=96$, $T_\chi/m_\chi=1$, and chemical potential $\mu_\chi/m_\chi=0$. The upper panel shows the absolute raw and projected KineticXGPU collision terms, evaluated in double precision with the analytical reduced kernel. For an exact Bose--Einstein distribution the continuum collision operator vanishes by detailed balance. In KineticXGPU, the off-grid occupation is interpolated through $y=\log[f/(1+f)]$ (cf. eq.~\eqref{eq:app_log_interp}), which is exactly linear in energy for a Bose--Einstein distribution. The raw collision term is therefore suppressed to the numerical round-off level before projection; the lower panel shows the smaller projection correction.
}
\label{fig:app_BE_fixed_point}
\end{figure}

As an independent check of the thermal fixed point, we also evaluate the quantum collision operator for an exact Bose--Einstein distribution, $f_{\rm BE}(E)
=
(\exp[(E-\mu_\chi)/T_\chi]-1)^{-1}.$ For this distribution, detailed balance requires $C_{\rm el}[f_{\rm BE}]=0$. As shown in figure~\ref{fig:app_BE_fixed_point}, the raw KineticXGPU collision term is suppressed to the numerical round-off level before the conservation projection is applied, while the projected result remains nearly indistinguishable from it.

\begin{figure}[H]
    \centering
    \includegraphics[width=0.93\textwidth]
    {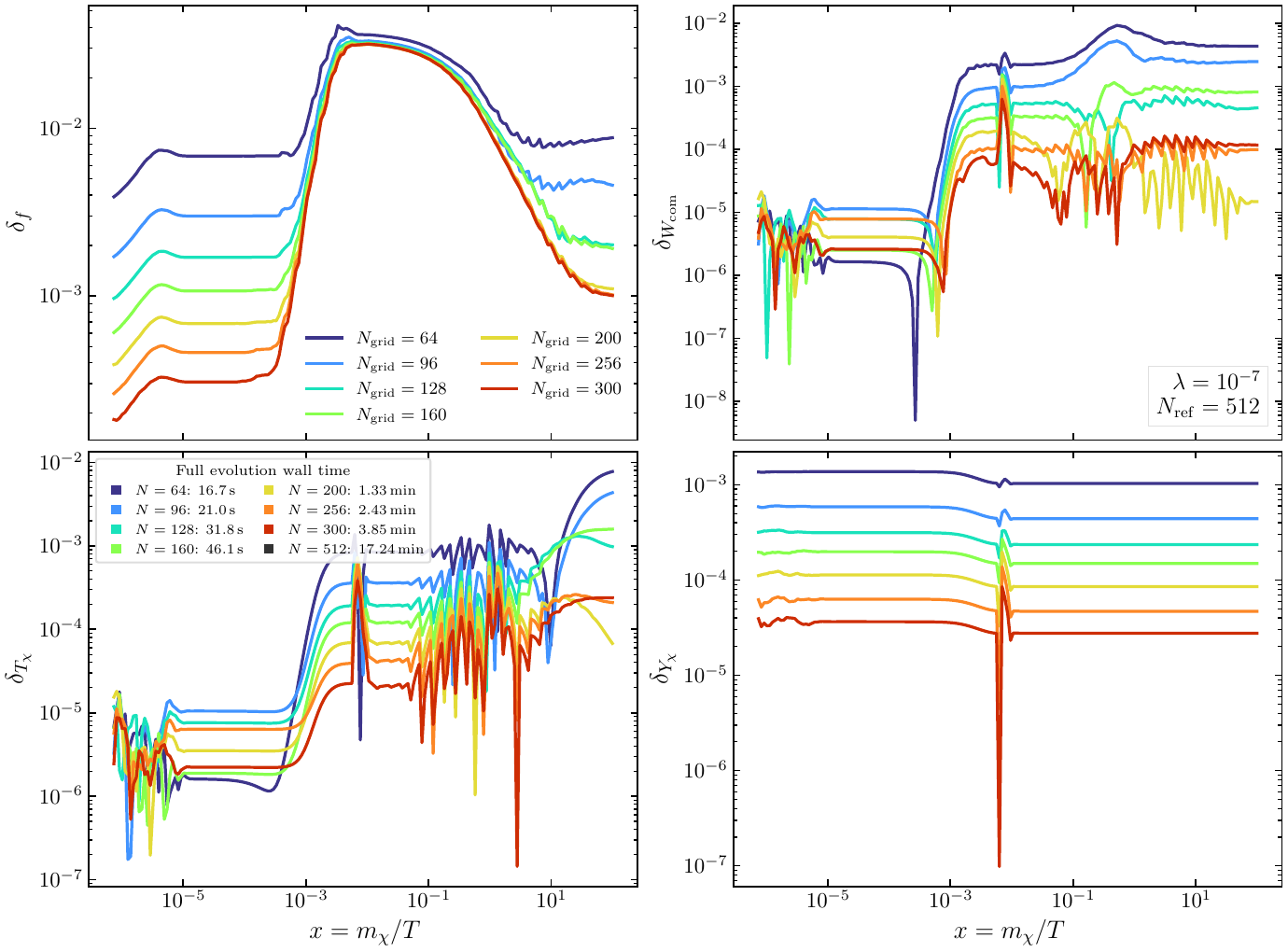}
    \caption{
    Momentum-grid convergence of the full cosmological evolution for the $100\,\mathrm{keV}$ benchmark with $\lambda=10^{-7}$. The $N_{\rm grid}=512$ solution is used as a finite-resolution reference. The panels show the distribution-level difference $\delta_f$ and the relative differences $\delta_{W_{\rm com}}$, $\delta_{T_\chi}$, and $\delta_{Y_\chi}$. The analytical reduced collision kernel and conservation projection are used at every resolution, in double precision on the GPU of table~\ref{tab:hardware}. At the production resolution $N_{\rm grid}=200$, the maximum deviations are $3.17\%$, $0.13\%$, $0.07\%$, and $0.02\%$, respectively. The inset key reports the end-to-end wall time of each evolution, excluding the preceding batch-size tuning.
    }
\label{fig:app_evolution_ngrid_convergence}
\end{figure}

Finally, we test the momentum-grid dependence of the full cosmological evolution. We use the benchmark of section~\ref{sec:results} with $\lambda=10^{-7}$ and repeat the evolution for $N_{\rm grid}=64$, 96, 128, 160, 200, 256, 300, and 512 with the batch sizes $B=64,96,48,160,96,96,32,$ and $24$, respectively. The analytical kernel and the conservation projection are enabled in every run, and all evolutions have the same adaptive time-integration settings. For the self-scattering regime, we use the Heun integrator described in section~\ref{subsec:time_step_integration} with relative and absolute tolerances $\mathrm{rtol}=10^{-2}$ and $\mathrm{atol}=2\times10^{-8}$, and an initial step size $\Delta u=10^{-4}$. The step-size safety factor is $0.9$. We take the $N_{\rm grid}=512$ solution as a finite-resolution reference and compare the full normalized distribution, the comoving warmness, the effective temperature, and the yield. For the distribution we define
\begin{equation}
    \delta_f(x)
    =
    \frac{
    \int d\ln q\,
    \left|P_{N_{\rm grid}}(q,x)-P_{512}(q,x)\right|
    }{
    \int d\ln q\,P_{512}(q,x)
    },
    \qquad
    P(q,x)\equiv \frac{1}{N}\frac{dN}{d\ln q},
\end{equation}
while for the integrated quantities we define
$\delta_Q=|Q_{N_{\rm grid}}/Q_{512}-1|$, with $Q=W_{\rm com}$, $T_\chi$, and $Y_\chi$.

The refinement study in figure~\ref{fig:app_evolution_ngrid_convergence} includes the condensate-injection epoch and finds a maximum distribution-level difference of $3.17\%$ between $N_{\rm grid}=200$ and 512. Over the same evolution, the maximum deviations in $W_{\rm com}$, $T_\chi$, and $Y_\chi$ are $0.13\%$, $0.07\%$, and $0.02\%$, respectively. The detailed phase-space distribution therefore converges more slowly than its low moments, as expected for a quantity sensitive to local momentum-space structure, while the physical observables reported in section~\ref{sec:results} remain stable under further momentum-grid refinement. The transient increase in $\delta_f$ occurs during the condensate-injection and relaxation epoch, when the narrow moving source is most sensitive to the finite momentum-grid resolution. After injection ceases, the discrepancy initially persists but subsequently decreases as elastic self-scattering smooths the local momentum-space structure and drives the different grid resolutions toward the same distribution.

\begin{figure}[t!]
    \centering
    \includegraphics[width=0.95\textwidth]
    {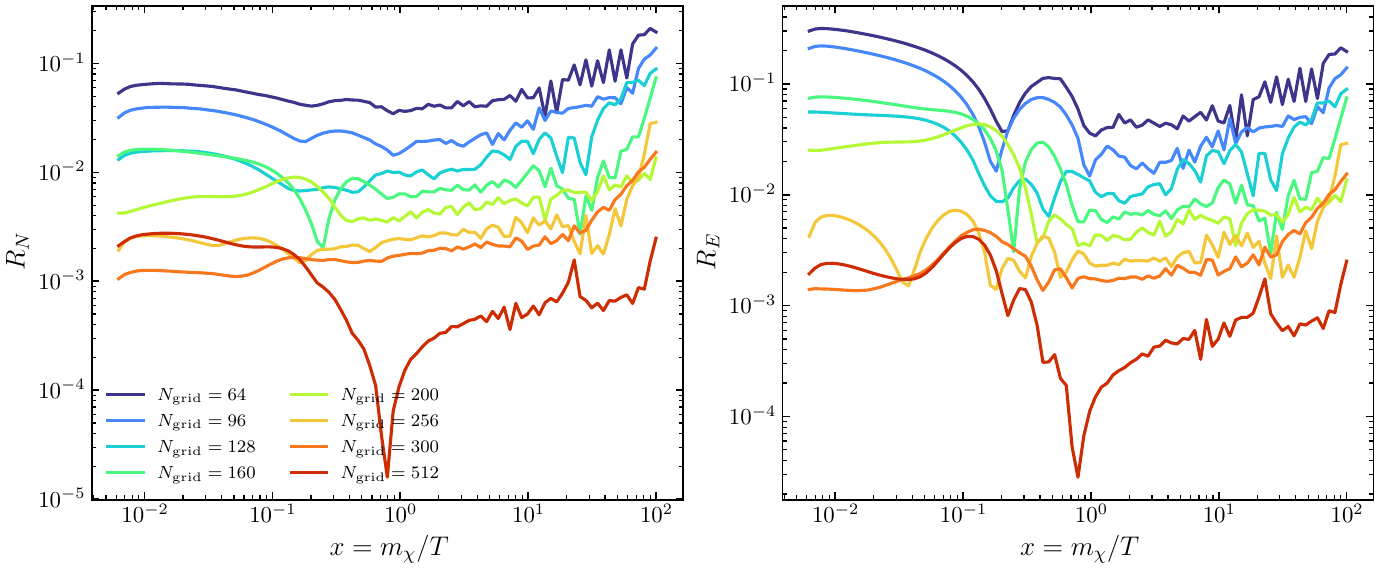}
    \caption{
    Fractional size of the number- and energy-weighted conservation corrections over the $100\,\mathrm{keV}$ evolution with $\lambda=10^{-7}$. At $N_{\rm grid}=200$, the maximum values are $R_N=1.36\%$ and $R_E=4.34\%$; for $N_{\rm grid}=512$, the corresponding maxima are $0.28\%$ and $0.42\%$. The projected operator has vanishing discrete number and energy moments at every resolution.
    }
\label{fig:app_projection_convergence_vs_x}
\end{figure}

We also follow the relative projection corrections $R_N$ and $R_E$ defined in eq.~\eqref{eq:app_projection_relative_correction} throughout the cosmological evolution shown in figure~\ref{fig:app_projection_convergence_vs_x}.

\section{Code availability and reproducibility}
\label{app:code}

\begin{table}[t]
    \centering
    \begin{tabular}{@{}ll@{}}
        \toprule
        Component & Specification \\
        \midrule
        System & Lenovo ThinkPad P15 Gen 1 \\
        CPU & Intel Core i7-10750H @ 2.60 GHz, 6 cores / 12 threads \\
        GPU & NVIDIA Quadro T2000 Mobile / Max-Q, 4 GB memory \\
        System memory & 32 GB RAM \\
        CUDA version & 12.2 \\
        \bottomrule
    \end{tabular}
    \caption{Hardware configuration used for the timing measurements shown in figure~\ref{fig:run_time}.}
    \label{tab:hardware}
\end{table}

The code is publicly available at \url{https://github.com/EsauCervantes/KineticXGPU}.  The repository contains the implementation of the self-collision operator, the hybrid freeze-in/self-scattering solver, the coupled Boltzmann-equation solver used for comparison, benchmark utilities, and the plotting routines used for the figures in this paper. The README provides installation instructions, examples of the Python package interface and command-line usage, and the scripts needed to reproduce the benchmark and cosmological runs.

The kernel integral, collision operators, and conservation projection are implemented in \texttt{collision.py} through \texttt{F\_contact()} and \texttt{F\_analytical()} for the contact-interaction kernel, \texttt{C\_MB()} for the Maxwell--Boltzmann collision term, \texttt{C\_quantum()} for quantum statistics, and \texttt{project\_self\_zero\_moments()}. The time integrator is implemented as \texttt{integrate\_rk4\_loga()} in \texttt{solver.py}, while the hybrid freeze-in/self-scattering evolution is defined in \texttt{run\_hybrid()}. The coupled Boltzmann-equation solver for eq.~\eqref{eq:cBE} is provided in \texttt{cBE\_solver.py}. The hardware details used for the timing measurements are shown in table~\ref{tab:hardware}.
\FloatBarrier


\bibliographystyle{JHEP}
\bibliography{biblio.bib}


\end{document}